\documentclass{revtex4}
\usepackage{graphicx,amsmath,amssymb,mathrsfs}
\usepackage{epsfig}

\begin{document}
\title{Environmental Superstatistics}
\author{ G.Cigdem Yalcin$^{1}$\footnote{
corresponding author, phone +90 212 455 57 00 ext 15489,
Fax +90 212 455 5855, gcyalcin@istanbul.edu.tr}
 and Christian Beck$^{2}$}
\affiliation{$^{1}$Department of Physics, Istanbul University,  34134,  Vezneciler, Istanbul, Turkey
\\
$^{2}$School of Mathematical Sciences, Queen Mary, University of London, London E1 4NS, UK}

\begin{abstract}
A thermodynamic device placed outdoors, or a local ecosystem, is subject to a variety
of different temperatures given by short-term (daily)
and long-term (seasonal) variations.
 In the long term a superstatistical description
makes sense, with a suitable distribution
function $f(\beta)$ of inverse temperature $\beta$ over which
ordinary statistical mechanics is averaged.
We show that $f(\beta)$
is very different
at different geographic locations, and typically exhibits
a double-peak structure for long-term data. For some of our data sets we also find a systematic
drift due to global warming. {For a simple superstatistical model system we show that the response
to global warming is stronger if temperature fluctuations are taken into account}.
\end{abstract}

\vspace{2cm}

\maketitle

\section{Introduction}

In nonequilibrium statistical mechanics, the
superstatistics technique
\cite{beck-cohen} is a powerful tool
to describe a large variety of complex systems
for which there is a change of environmental conditions and temperature fluctuations on a large scale
\cite{swinney,touchette,souza,chavanis,jizba,
frank,celia,straeten,hanel}.
A superstatistical
complex system is mathematically described as a
superposition of
several statistics,
one corresponding to local equilibrium statistical mechanics
and the other one corresponding to a slowly
varying parameter $\beta$ of the system.
Essential for this approach is the fact that there is sufficient time scale separation,
i.e. the local relaxation time of the system must be much shorter than the
typical time scale on which $\beta$ changes.
Many interesting applications
of the superstatistics concept have been worked out for a variety
of complex systems,
for example the analysis of
train delay statistics\cite{briggs},
hydrodynamic turbulence \cite{prl}, cancer survival statistics
\cite{chen} and some other applications as well, see
\cite{daniels,soby,dixit,abul-magd,rapisarda,cosmic}.

In this paper we want to deal with environmental aspects
of superstatistics, in the sense that we ask what distributions
of inverse temperatures are seen by thermodynamic devices that are
kept in open air outside a constant-temperature environment.
Clearly this question is technically very relevant as many devices need
to operate under strong temperature fluctuations,
as given by  either daily temperature
fluctuations, or seasonal variations, or even long-term climatic changes. 
{Besides thermodynamic devices, one may also be interested
in local complex ecosystems (such as e.g. biological populations) coupled to a
changing temperature environment.} 
We will analyse in detail
inverse temperature distributions at various geographic locations.
These environmentally important distributions are different from
standard examples of distribution functions discussed so far
in the literature,
such as the $\chi$-square, inverse $\chi$-square or lognormal distribution
\cite{swinney}.
As it is well-known, superstatistics based on $\chi$-square distributions leads to
$q$-statistics \cite{tsallis1,tsallis-book}, whereas other distributions lead to
something more complicated.
A major difference is that the environmentally observed probability
densities of inverse temperature typically exhibit a double-peak structure, thus requiring
a different type of superstatistics than what has been done so far.
In the original setting described in \cite{beck-cohen},
the analysis was based on single-peaked distributions with a sharp
maximum \cite{beck-cohen}. For environmentally relevant
temperature distributions, this concept has to be broadened.

{A central point for the applicability of the superstatistics concept is
the existence of suitable time scale separation of the dynamical evolution, or more
generally the existence of a hierarchy of time scales which are well
separated. In the simplest case this just means there
are two different time scales such that the typical variation of $\beta$ takes place
on a much larger time scale than the local relaxation time of the system that is influenced
by the given temperature environment. For meteorological and climatic systems
there is indeed  a hierarchy of different relevant time scales. It starts with time scales due to
the turbulent dynamics of the air which are well below the daily temperature oscillations;
also there are synoptic meteorological disturbances of the order of about 2 days.
We arrive at seasonal variations (circular statistics) and at larger scales
at inter-annual variability. Finally, at largest scales there are long-term trends due
to climatic and geological changes. Our aim in this paper is not a
detailed analysis of all these various dynamical phenomena on various time scales
(see, e.g. \cite{ghil,ghil2,allen,porporato2,kantz} and references therein),
but the application of superstatistical techniques, given the fact that there
is time scale separation. This means effectively
we consider a given subsystem or device (which may be a technical
thermodynamic device, but in a more general setting also a
given local ecosystem depending on temperature and precipitation dynamics)
and then look at a generalized statistical mechanics description of it.
For this it is necessary to fully understand the statistics of (inverse) temperature fluctuations---
either as sampled over long-term records or conditioned on particular periods
(say summer or winter). While in principle one could also do
superstatistics on shorter (turbulent) time scales of air movement,
in this paper we restrict ourselves to the long-term statistical properties.

This paper is organized as follows. In section 2 we look at monthly data
(essentially eliminating seasonal variations) at various geographical locations,
and check how well the data are described by Gaussian distributions.
In section 3 we look at long-term data including seasonal variations, which
induce double-peaked distributions, but with specific differences at
different geographical locations, depending on local climate.
In section 4 the results are interpreted in terms of the K\"oppen-Geiger climate
classification scheme. In section 5 we discuss why the superstatistics
relevant for environmental temperature fluctuations is different from what has
been done so far for single-peaked distributions. In section 6 we
deal with global warming, and the interesting effects that occur due to fluctuations
in superstatistical models if a parameter such as the mean of the temperature distributions increases
slightly. Finally, our conclusions are summarized
in section 7.}

\section{Observed inverse temperature distributions---monthly data}

When looking of temperature distributions as given by real data,
one clearly has to specify the relevant time scale first. Short-term
temperature distributions (dominated by daily fluctuations)
are very different from long-term data (dominated by seasonal
variations). For very long data records, one also has to take into
account non-stationary behaviour
due to climate change.

We start with short-term data.  Fig.~1, 2 and 3 show as
an example a time series of hourly  measured surface inverse temperatures in Ottawa during November 2011,
as well as in Vancouver during May 2011 and December 2011.

The frequency pattern is dominated by 1 oscillation per 24h, corresponding to
the day-night temperature differences. Superimposed to this are stochastic fluctuations
due to different weather conditions and turbulent fluctuations. What is also clearly visible in the figures are
small systematic trends. In Fig.~1 (November in Ottawa) the average inverse temperature $\beta$ is slightly increasing
over the month of November,
since the temperature slightly decreases as winter is approaching. In Fig.~2 (May in Vancouver) the average
$\beta$ is slightly decreasing, since summer is approaching. No clear systematic trend is visible
for the December data of Fig.~3.

\begin{figure} [ht]
\epsfig {file=2011nov-ottawa-data.eps, angle=0., width=11cm}
\caption{Time series of hourly measured inverse temperature in Ottawa during November 2011}

\vspace{2cm}

\epsfig {file=2011may-vancouver-data.eps, angle=0., width=11cm}
\caption{Time series of hourly measured inverse temperature in Vancouver during May 2011}
\end{figure}

\clearpage

\begin{figure} [ht]
\epsfig {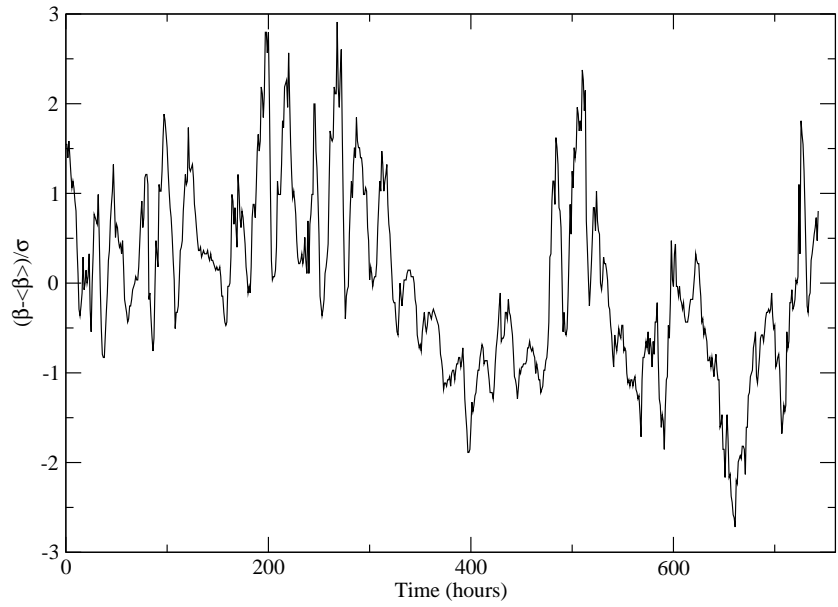}
\caption{Time series of hourly measured inverse temperature in Vancouver during December 2011}
\end{figure}

\vspace{1cm}

The (inverse) temperature time series for a single month is approximately Gaussian distributed.
This is illustrated in Figs.~4-6,
which shows the histograms of the example time series of Figs~1-3.

\vspace{2cm}

\begin{figure} [ht]
\epsfig {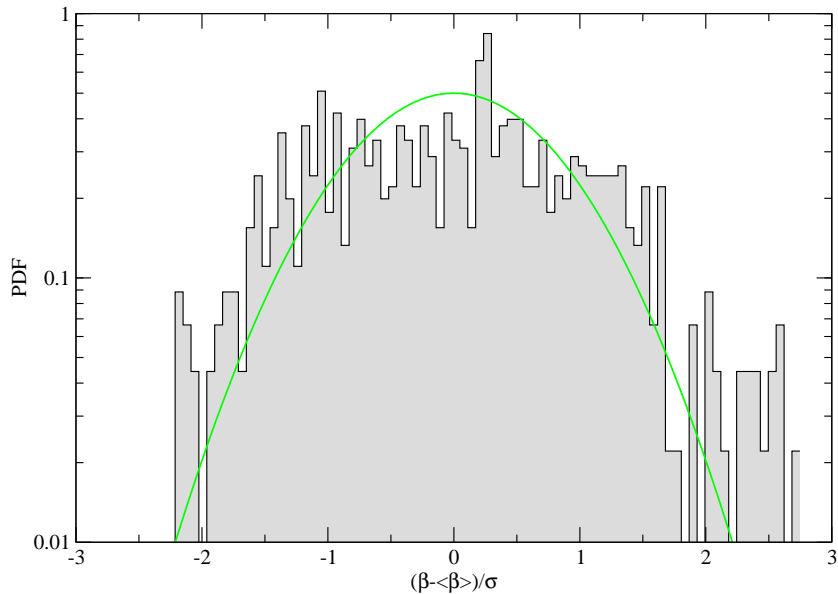}
\caption{Gaussian fit to the histogram of the hourly measured inverse temperature in Ottawa during November 2011}
\end{figure}

\clearpage

\begin{figure} [ht]
\epsfig {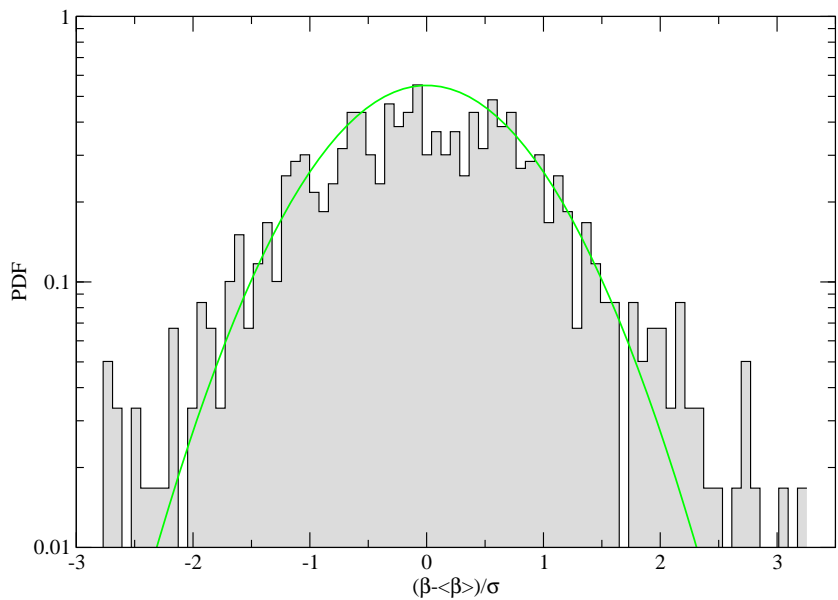}
\caption{Gaussian fit to the histogram of the hourly measured inverse temperature in Vancouver during May 2011}
\end{figure}

\vspace{2cm}

\begin{figure} [ht]
\epsfig {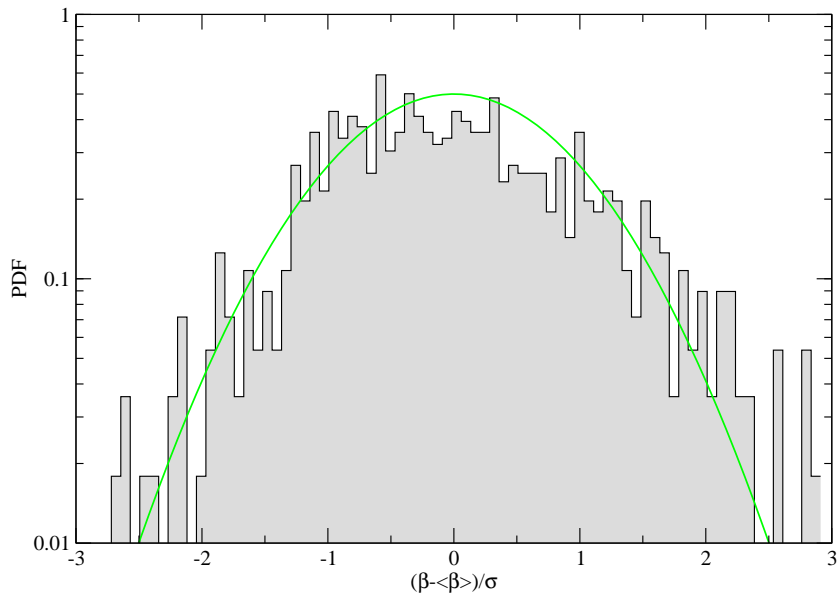}
\caption{Gaussian fit to the histogram of the hourly measured inverse temperature in Vancouver during December 2011}
\end{figure}

{Consider now a thermodynamic device with energy levels $E_i$ that is kept outside and hence
subject to slowly varying inverse temperatures. Our interpretation of `thermodynamic device'
is very general here. One may also think of an ecosystem or other complex system (including
biological populations) that is influenced or in fact, dependent, on the temperature (and precipitation) of its
environment. Suppose that locally, for constant inverse temperature, the device is properly described by
statistical mechanics with ordinary Boltzmann factors $e^{-\beta E}$. Assuming time scale separation,
so that the system can quickly enough relax to local equilibrium, the long-term behaviour is
properly described by a mixture of Boltzmann factors with different temperatures, weighted
with a function $f(\beta)$ that describes how often a given inverse temperature of the
environment is observed. This is the realm of superstatistics \cite{beck-cohen}.}
The effective statistical mechanics of this thermodynamic device (or ecosystem) is indeed
described by effective Boltzmann factors $B(E)$ of the form
\begin{equation}
B(E)=\int_0^\infty f(\beta) e^{-\beta E} d\beta \label{effe}
\end{equation}
{In our case, for monthly distributions as displayed in
 Fig. 4-6,  $f(\beta)$ is indeed a single-humped probability density around the mean inverse temperature
that month, in good approximation given by a Gaussian distribution}.
The integration over $\beta$ yields non-Boltzmannian distributions $B(E)$, whose details depend
on $f(\beta)$. So for example the marginal distribution of velocities of an ideal gas for which $\beta$
fluctuates is not Gaussian anymore, due to the integration over $\beta$ in eq.~(\ref{effe}).

Better statistics is obtained if we sample the data of a given month over many years.
      We sampled hourly measured inverse temperature time series, restricted to the month of November,
      over the period 1966-2011 for Ottawa.
      Fig.~7 shows this time series (restricted to November months and joined together).
The corresponding histogram is well-fitted by a Gaussian (Fig.~8).

{However, let us remark that Gaussian distributions in (inverse) temperature can of course only be an approximation.
They cannot be exact since in principle
they allow for negative $\beta$, whereas $\beta$ should always be positive.
A better model would, for example, be a $\chi$-square distribution in $\beta$, or a lognormal
distribution, which always has positive $\beta$. Close to their maxima these distributions
still look similar to Gaussians. We should remark at this point that temperature distributions
at various locations on earth have of course been analysed before. A recent paper \cite{ruff} provides evidence
that the distributions (when seasonal fluctuations are eliminated) are sometimes Gaussian, sometimes
non-Gaussian, depending on location. While for our data in typical
cases we see near-Gaussian behavior,
in \cite{ruff} evidence for power-law distributions and exponential tails at some special locations has been provided.}

\vspace{2cm}

\begin{figure} [ht]
\epsfig {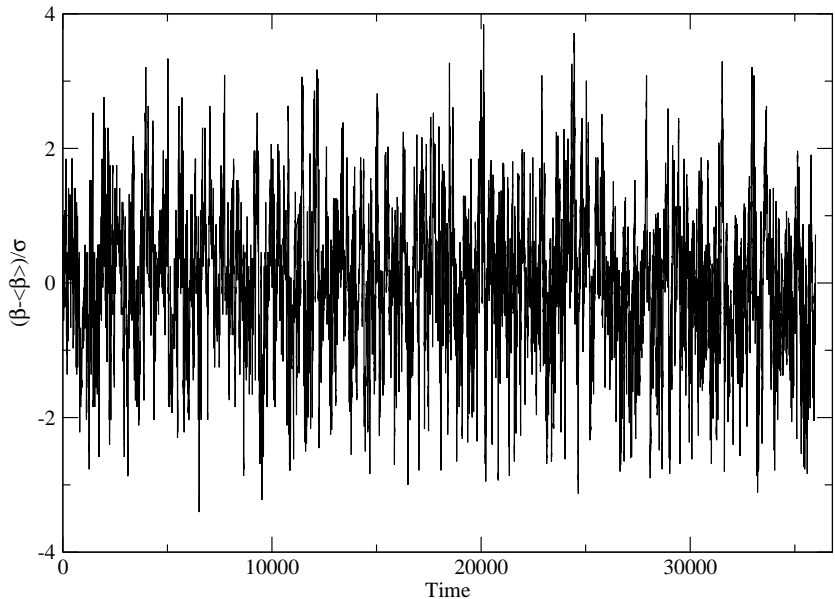}
\caption{Time series of hourly measured inverse temperature in Ottawa for November, sampled from 1962 to 2011 for 50 years}
\end{figure}

\vspace{2cm}

\begin{figure} [ht]
\epsfig {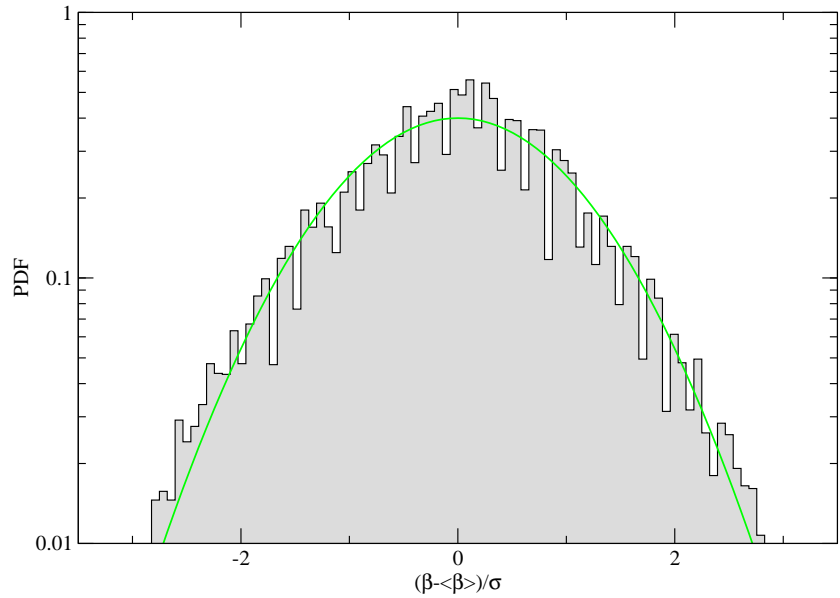}
\caption{Gaussian fit to the histogram of the hourly measured inverse temperature in Ottawa for 50 November months}
\end{figure}

\vspace{2cm}

What is generally interesting in our data is that there is a systematic trend of a changing mean
on long time scales. Close inspection of Fig.~7
shows already to the bare eye that the average $\beta$ is slightly decreasing over the years,
meaning the average November temperature has slightly
increased in Ottawa over the past 50 years. Clearly, this effect is related to global warming,
here measured at a particular location.
The same trend of a slightly decreasing average $\beta$ is also visible for
the May-data of Vancouver (Fig.~9), which overall are also well fitted by a Gaussian
distribution (Fig.~10). Slightly more volatile behaviour is observed in December (Fig.~11),
with some outbursts to large $\beta$ (some very cold days) occurring relatively frequently, see Fig.~12.

{In line with the observations in \cite{ruff}, we see for particular seasons and particular
locations deviations from Gaussian behavior, although this seems
somewhat exceptional. For example, Figure 12 (winter in Vancouver)
is an example where there seems to be an exponential tail for large values of $\beta$.}

\vspace{2cm}

\begin{figure} [ht]
\epsfig {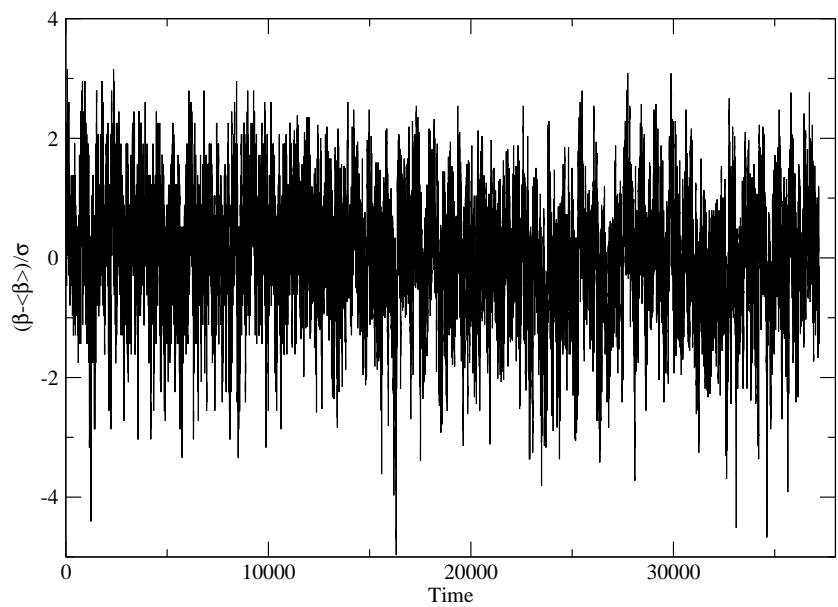}
\caption{Time series of hourly measured inverse temperature in Vancouver for 50 months of May}

\end{figure}

\vspace{2cm}

\begin{figure} [ht]
\epsfig {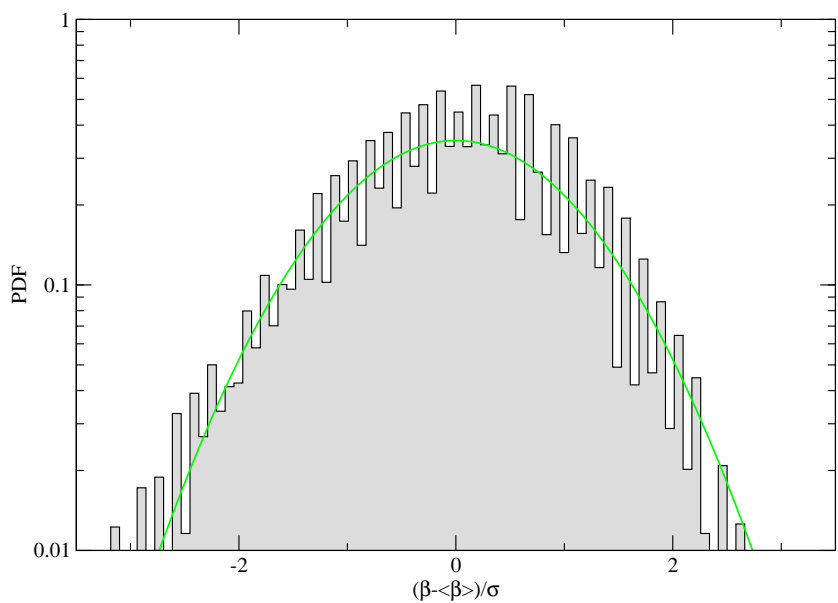}
\caption{Histogram the hourly measured inverse temperatures in Vancouver for 50 months of May}
\end{figure}

\clearpage

\begin{figure} [ht]
\epsfig {file=50dec-vancouver-data.eps, angle=0., width=12cm}
\caption{Same as Fig.~9 but for December}

\vspace{2cm}

\epsfig {file=50dec-vancouver.eps, angle=0., width=12cm}
\caption{Same as Fig.~10 but for December}
\end{figure}

\clearpage

Typically, the distributions for most months are in reasonably good approximation Gaussian distributions.
They describe random-like fluctuations around the mean temperature in that month
with oscillating behaviour super-imposed due to day-night temperature
differences. In addition, on very long time scales there is
a small systematic drift of the mean temperature to larger values, due to
climate change. A thermodynamic device, to be used only in a particular
month of the year, would be typically described by a superstatistics where
the relevant inverse temperature distribution $f(\beta)$ is close to
a Gaussian, at least close to its maximum, and where the mean of that Gaussian (and possibly also
the variance) has a very small time dependence
over
a time scale of decades.

\section{Observed inverse temperature distributions--- yearly data}

We now deal with thermodynamic devices (or complex local ecosystems) that are kept outside during the entire year,
so that full seasonal variations are becoming relevant.
The observed probability distributions
become more complex, depending on the climatic conditions
at the particular location. Examples of time series that we investigated in somewhat more detail are
shown in Figs.~13-21, these are daily measured inverse mean temperatures in various locations sampled over
many years.
Clearly, the day-night oscillations are now less relevant, and what becomes dominating
are seasonal variations \cite{lu1,lu2,lu3}. The data now have
a dominating frequency corresponding to 1 oscillation per 365 days due to seasonal
variations, modulated by stochastic fluctuations due
to different weather conditions on a time scale of days, plus large-scale annual fluctuations
(some years warmer or colder than others) and systematic
trends due to climate change.

We have investigated time series data for 8 different locations in different climatic zones.
These are
Darwin (Northern Territory, Australia) (1975-2011) \cite{australia},
Santa Fe (New Mexico, USA) (1998-2011) \cite{ncdc},
Dubai (United Arab Emirates)(1974-2011) \cite{dubai},
Sydney (New South Wales, Australia) (1910-2011) \cite{australia},
Central England (London, Bristol, Lancashire) (1910-2011) \cite{uk1,uk2},
Vancouver (British Columbia, Canada) (1937-2011) \cite{canada},
Hong Kong (PRC) (1997-2011) \cite{hongkong},
Ottawa (Ontario, Canada) (1939-2011) \cite{canada},
and Eureka (Nunavut, Canada) (1951-2011) \cite{canada}, respectively.

These time series are heavily influenced
by seasonal variations and fluctuations around the typical yearly
behaviour.
Interesting enough, for some locations a systematic trend (a decrease in
the average inverse temperature over the years) is again visible
to the bare eye. The locations where such a systematic trend
is clearly seen are Dubai (Fig.~15), Sydney (Fig.~16),
Vancouver (Fig.~18), Ottawa (Fig.~20), and Eureka (Fig.~21).
Obviously these are local manifestations of the effects of global warming.

\vspace{2cm}

\begin{figure} [ht]
\epsfig {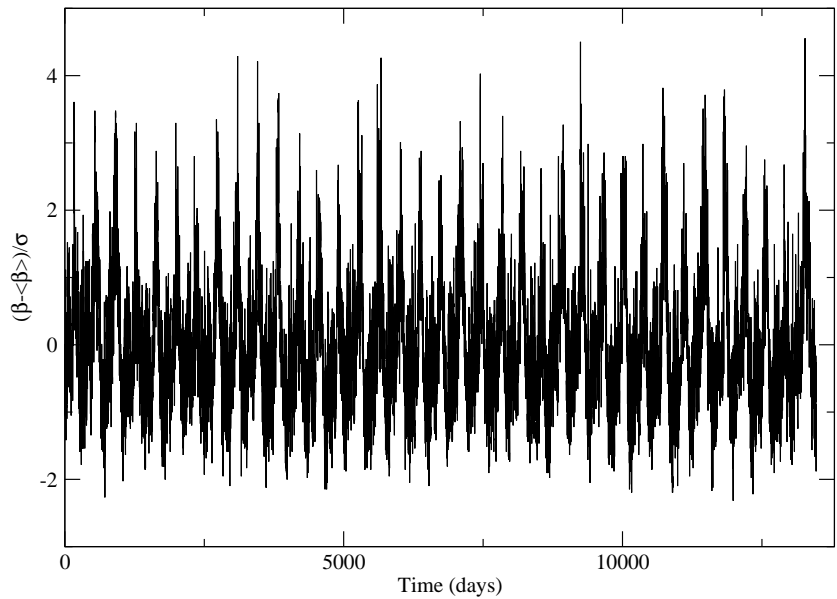}
\caption{Daily measured inverse mean temperature in Darwin 1975-2011}
\end{figure}

\clearpage

\begin{figure} [ht]
\epsfig {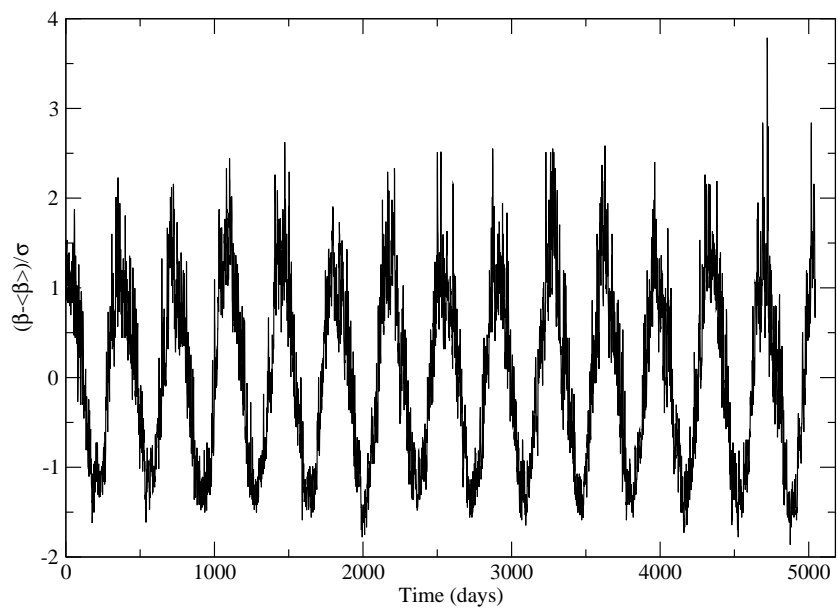}
\caption{Daily measured inverse mean temperature in Santa Fe 1998-2011}
\end{figure}

\vspace{2cm}


\begin{figure} [ht]
\epsfig {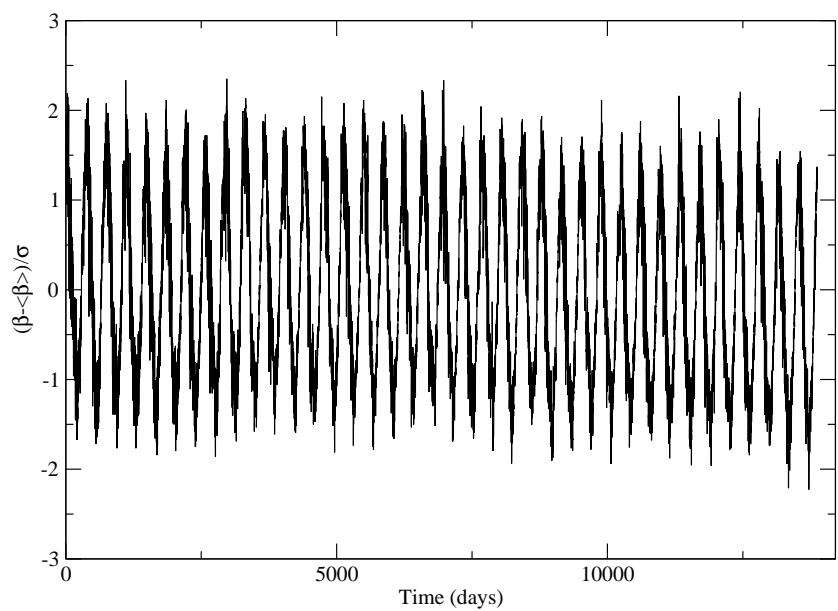}
\caption{Daily measured inverse mean temperature in Dubai 1974-2011}
\end{figure}

\clearpage

\begin{figure} [ht]
\epsfig {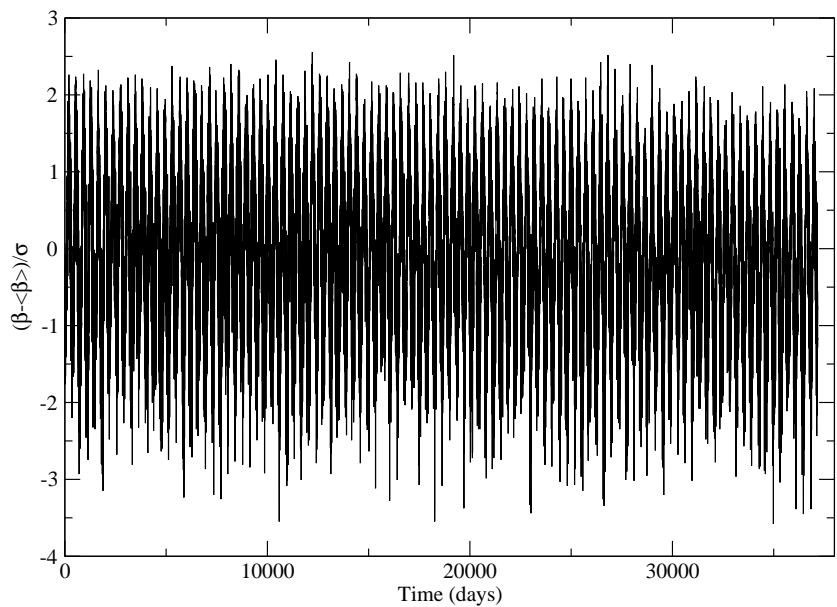}
\caption{Daily measured inverse mean temperature in Sydney 1910-2011}
\end{figure}

\vspace{2cm}


\begin{figure} [ht]
\epsfig {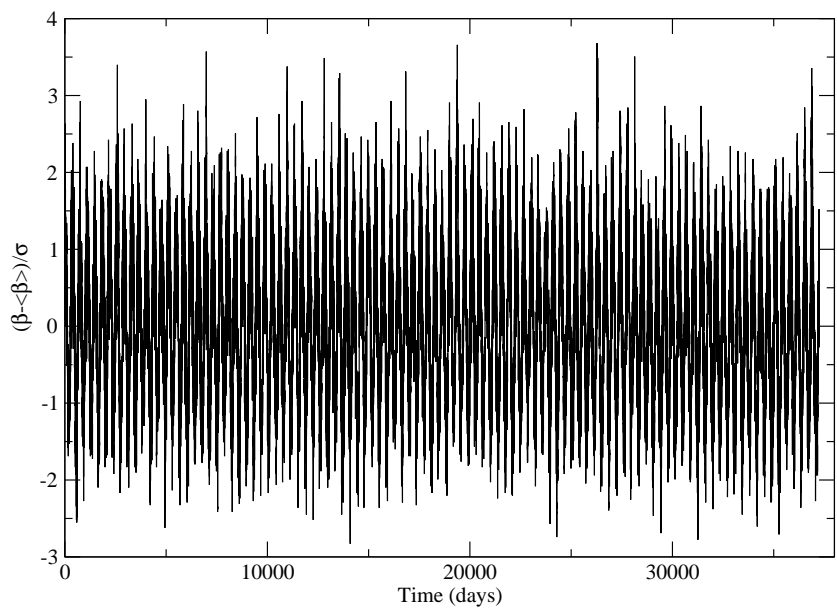}
\caption{Daily measured inverse mean temperature in Central England (Lancashire, London and Bristol) 1910-2011}
\end{figure}

\clearpage

\begin{figure} [ht]
\epsfig {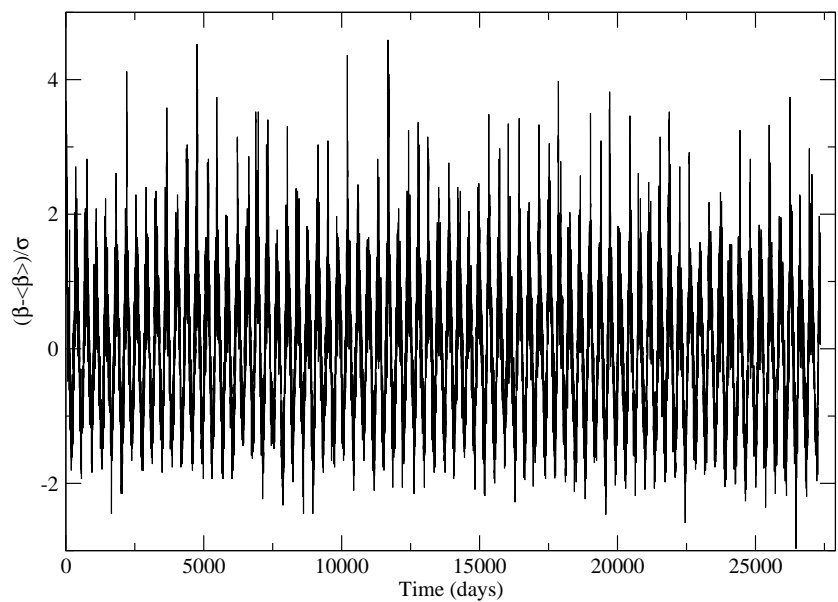}
\caption{Daily measured inverse mean temperature in Vancouver 1937-2011}
\end{figure}

\vspace{2cm}


\begin{figure} [ht]
\epsfig {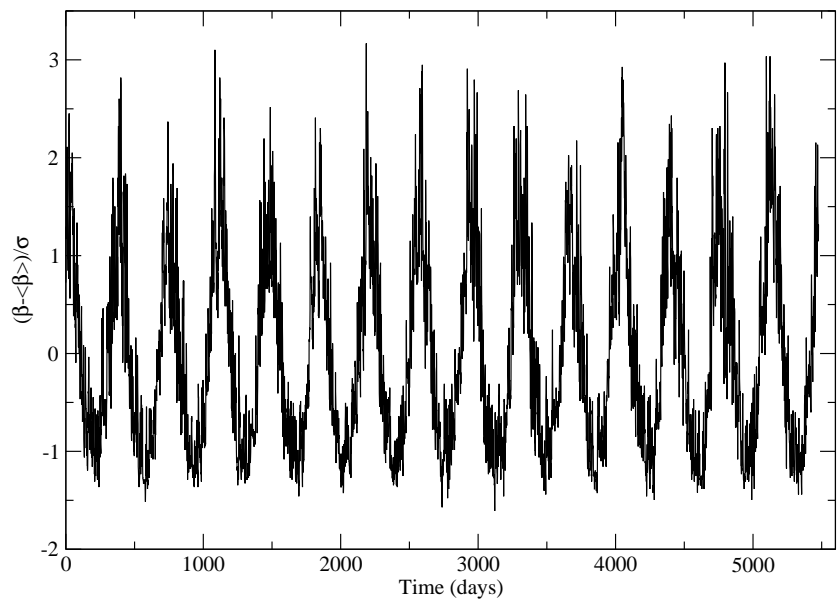}
\caption{Daily measured inverse mean temperature in Hong Kong 1997-2011}
\end{figure}

\clearpage

\begin{figure} [ht]
\epsfig {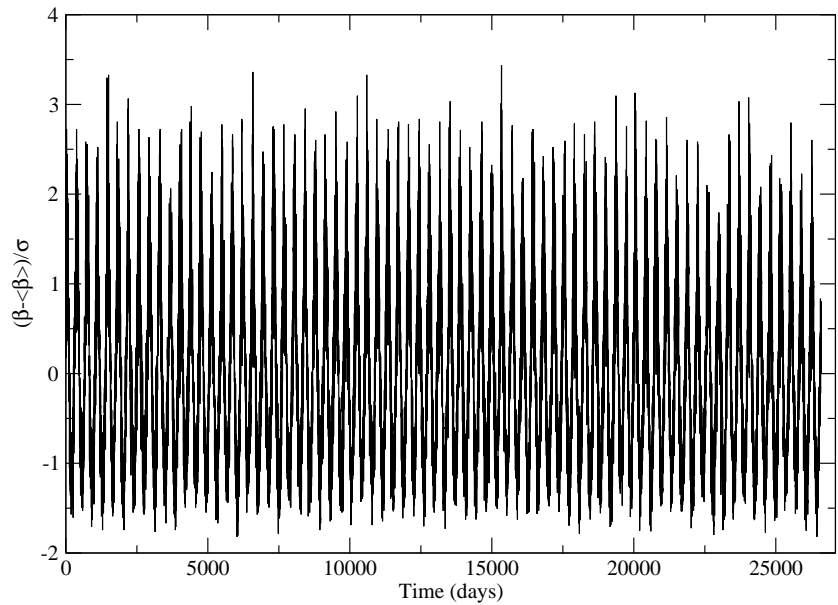}
\caption{Daily measured inverse mean temperature in Ottawa 1939-2011}
\end{figure}

\vspace{2cm}

\begin{figure} [ht]
\epsfig {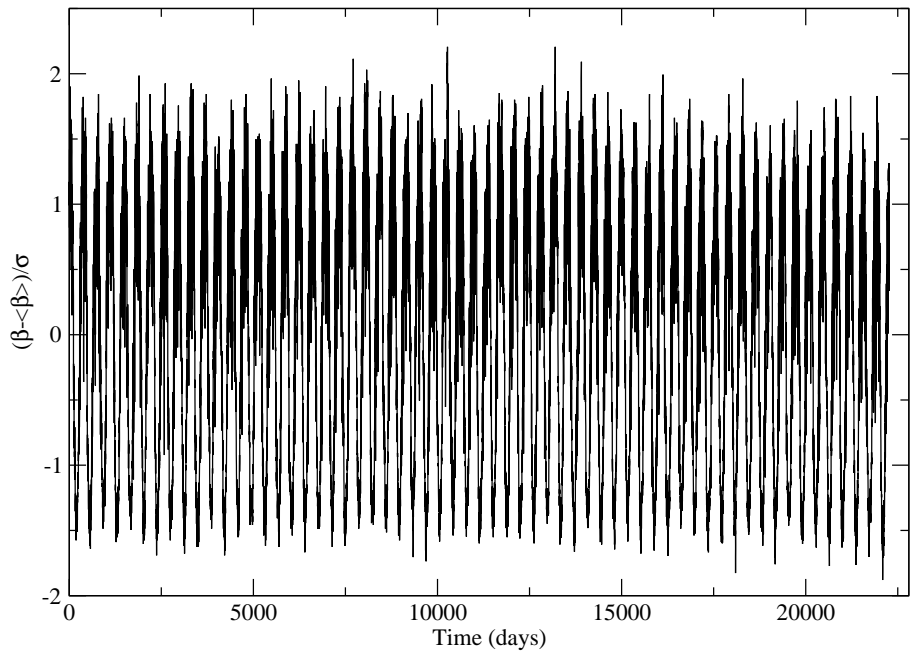}
\caption{Daily measured inverse mean temperature in Eureka 1951-2011}
\end{figure}

Figs.~22-30 show the histograms of inverse temperature time series for the various locations.
These distributions are the ones relevant for the superstatistical description of thermodynamic devices that
need to work in the open air and that are supposed to work properly for
continuous operation over many years. Similarly, these are also the distributions
seen by biological populations that don't have the luxury of artificial heating
or air-conditioning. The details of the distribution
of course heavily depend on the climate zone of the location.

\vspace{1cm}

A typical observation is now that the relevant distributions have a double-peak structure, at least
for non-tropical locations.  Broadly, the left peak (small $\beta$)
corresponds to summer and the right peak (large $\beta$) to winter.
The entire distribution can be very roughly regarded as a superposition of
two Gaussians, with some intermediate behaviour between the peaks.
This intermediate behaviour is more pronounced for geographical
locations that have big differences between summer and winter temperatures.

Apparently, for climates with strong differences of winter and summer temperatures the two peaks of our Gaussian fits (Fig.~22-30)
are far apart.
Table 1 lists the two temperatures where the two maxima in the histogram occur.
They are consistent with an average temperature observed during a
couple of months corresponding to summer and winter, respectively, at the different geographical locations.
{Some special locations, for example Darwin in Fig.~22, show again deviations from
Gaussian behavior for $f(\beta)$ , with exponential-like tails for large $\beta$, and sub-Gaussian behavior
for small $\beta$ (the red line in that figure is a Gaussian)}.

\begin {table}
\begin{center}
\begin{tabular}{|c|l|l|l|l|l|l|l|l|}
\hline
                $Location  \ and \ time \ period $ &$Summer (Celcius) $  &$Winter (Celcius) $  &$\beta_{1}$  &$\beta_{2}$ \\\hline

Darwin  (1975-2011)                  & \ \ \ \ \ \ \   28.32      &\ \ \ \ \ \ \      -           &\ \       1.8     \ \  &\ \          - \ \        \\\hline
Santa Fe (1998-2011)                &  \ \ \ \ \ \ \  22.06     &\ \ \ \ \ \ \ 3.31          &\ \        18     \ \  &\ \         4  \ \        \\\hline
Dubai (1974-2011)                     & \ \ \ \ \ \ \ 33.03        &\ \ \ \ \ \ \ 20.39       &\ \        7       \ \  &\ \         8  \ \         \\\hline
Sydney (1910-2011)                  &\ \ \ \ \ \ \  20.22        &\ \ \ \ \ \ \ 13.78       &\ \        2        \ \  &\ \        3.5 \ \        \\\hline
Central England (1910-2011)  & \ \ \ \ \ \ \ 14.26        &\ \ \ \ \ \ \ 7.12         &\ \        3.6     \ \  &\ \        1.4 \ \         \\\hline
Vancouver (1937-2011)            & \ \ \ \ \ \ \ 15.80        &\ \ \ \ \ \ \ 7.05         &\ \        5         \ \ &\ \          2  \ \          \\\hline
Hong Kong (1997-2011)           & \ \ \ \ \ \ \ 27.99        &\ \ \ \ \ \ \ 19.31       &\ \        10       \ \  &\ \        2.7\ \          \\\hline
Ottawa (1939-2011)                   & \ \ \ \ \ \ \  18.49       &\ \ \ \ \ \ \ -1.04        &\ \        12       \ \  &\ \         2.1\ \         \\\hline
Eureka (1951-2011)                   & \ \ \ \ \ \ \  4.21         &\ \ \ \ \ \ \ -36.3        &\ \        43       \ \  &\ \          5.5\ \         \\\hline

\end{tabular}
\caption{Maxima (in degree Celsius) and variance parameters $\beta_{1} $,~$\beta_{2}$
of the two Gaussians $\sim e^{-\beta_i (\beta-\bar{\beta})^2}$ used in the fits.
}
\end{center}
\end{table}


\vspace{2cm}

\begin{figure} [ht]
\epsfig {file=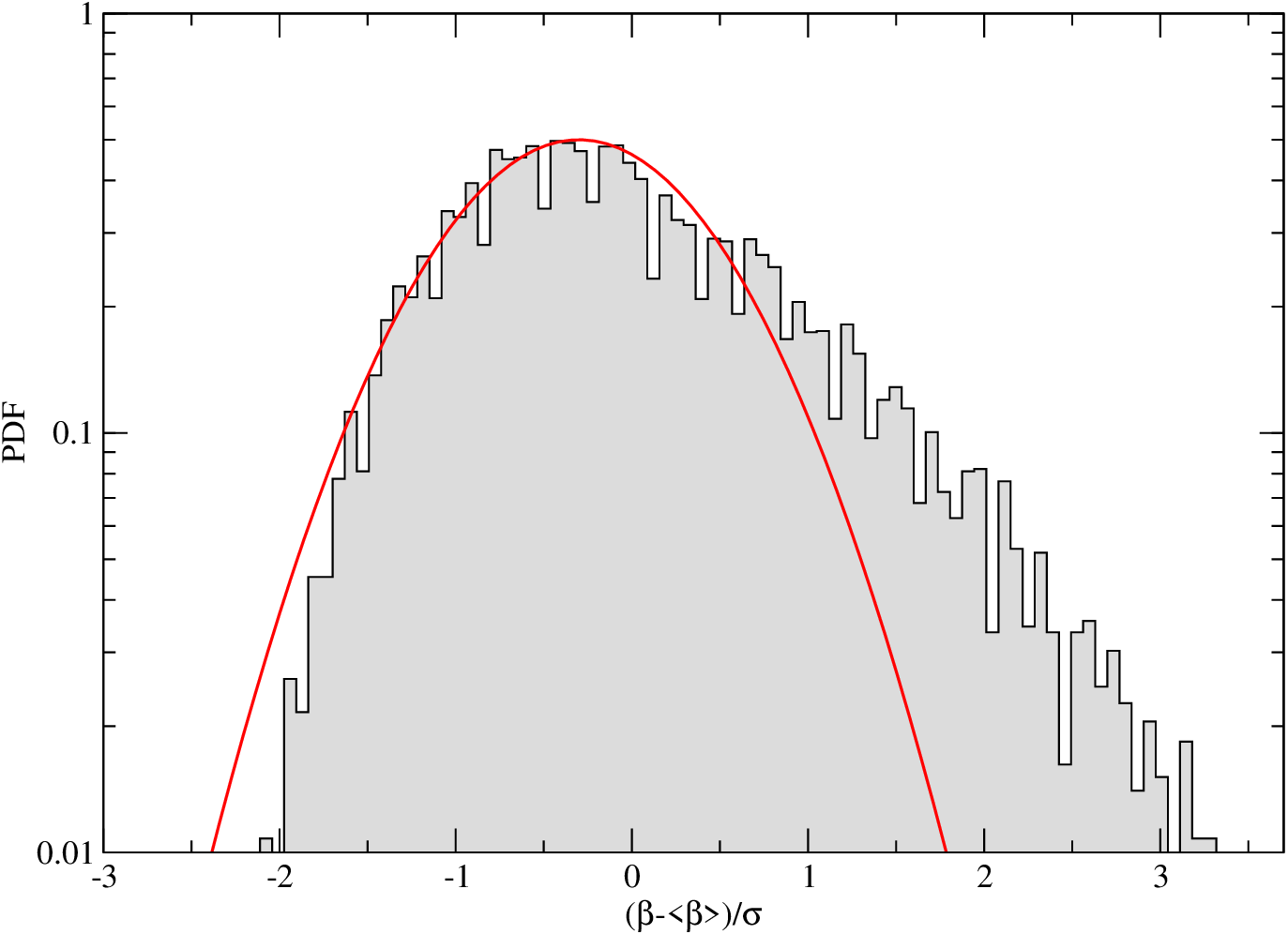, angle=0., width=11cm}
\caption{Distribution of the daily measured inverse mean temperature in Darwin for 1975-2011}
\end{figure}

\vspace{2cm}

\begin{figure} [ht]
\epsfig {file=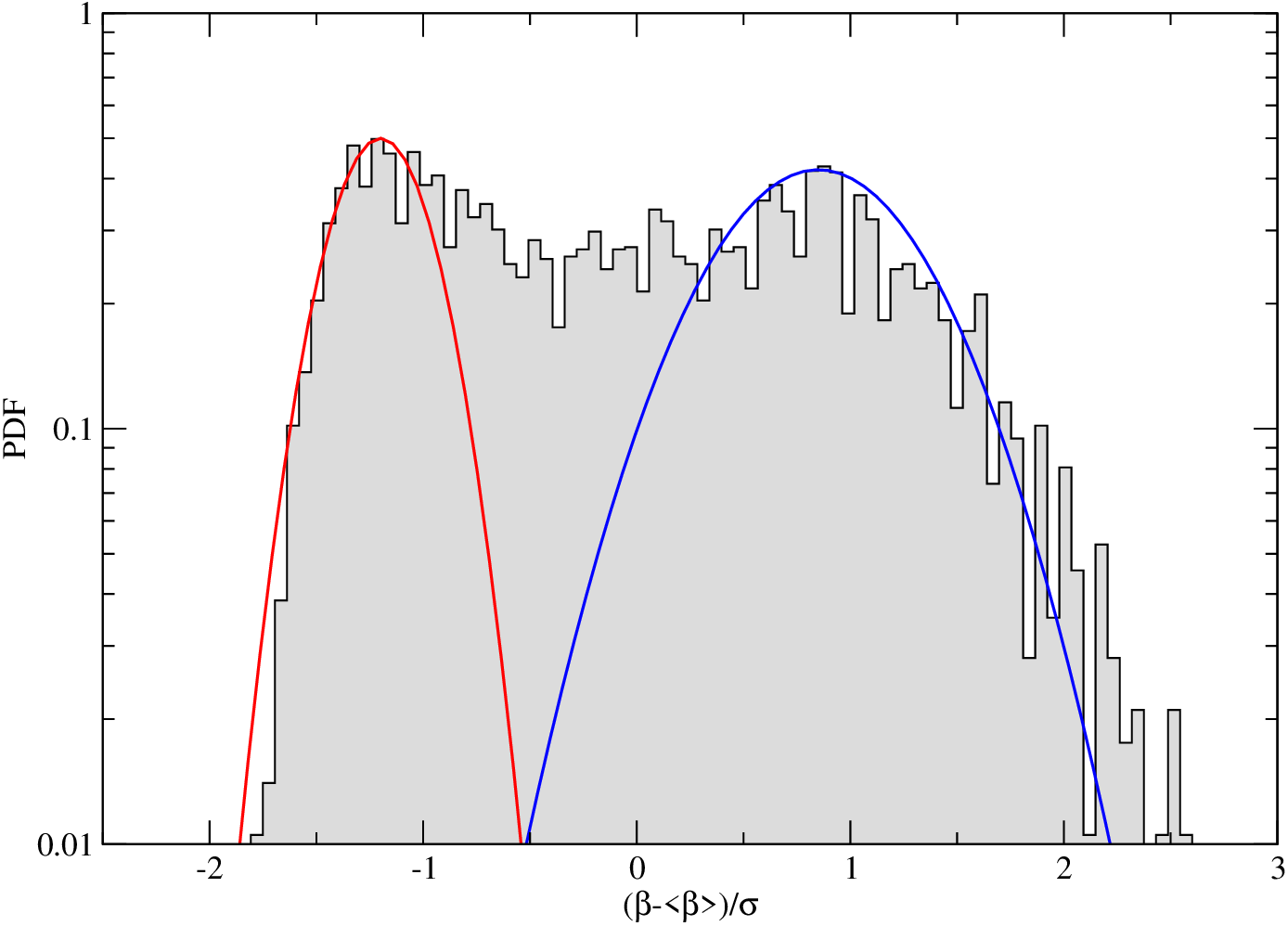, angle=0., width=11cm}
\caption{Distribution of the daily measured inverse mean temperature in Santa Fe for 1998-2011}

\vspace{2cm}

\epsfig {file=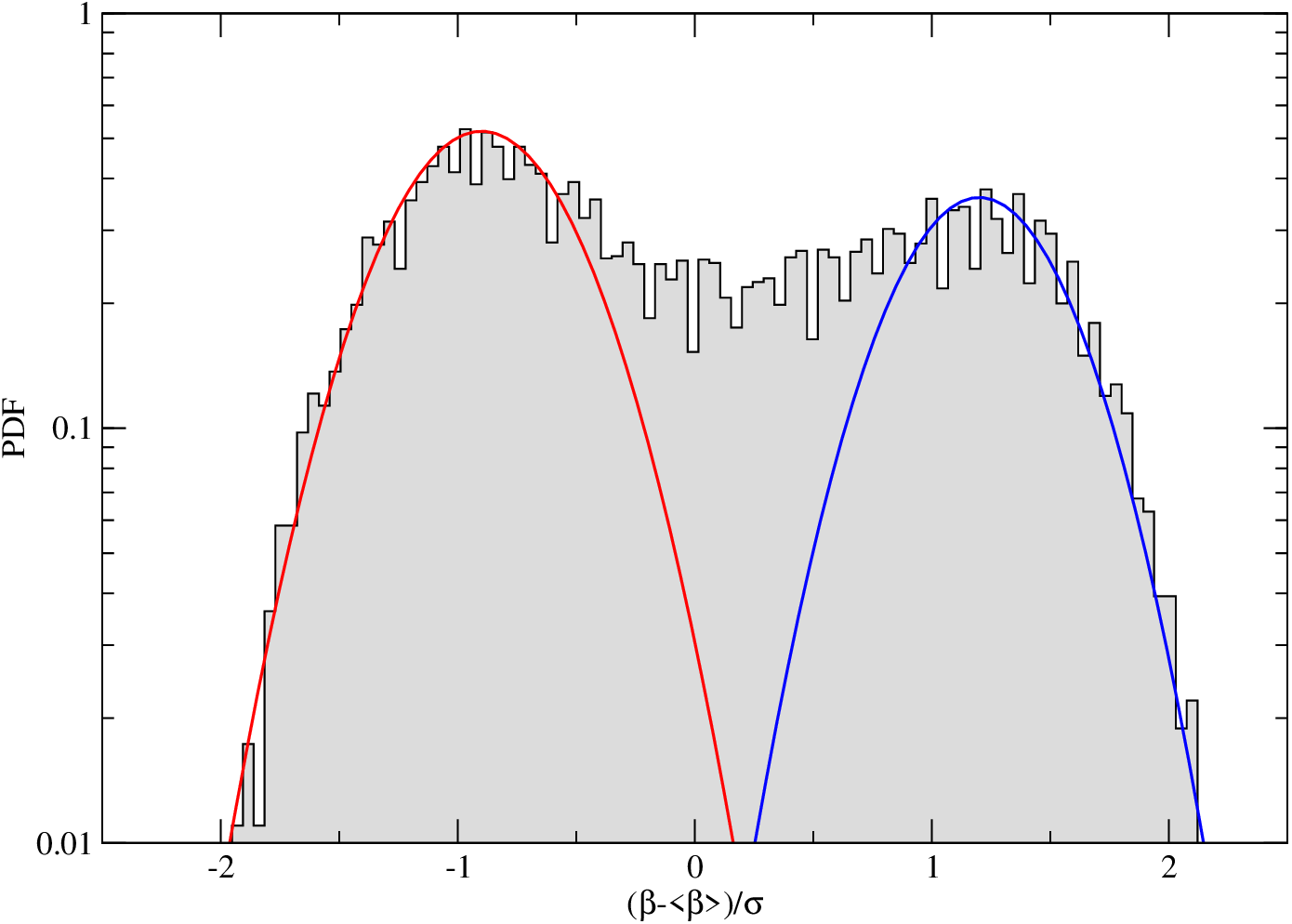, angle=0., width=11cm}
\caption{Distribution of the daily measured inverse mean temperature in Dubai for 1974-2011}
\end{figure}

\clearpage


\begin{figure} [ht]
\epsfig {file=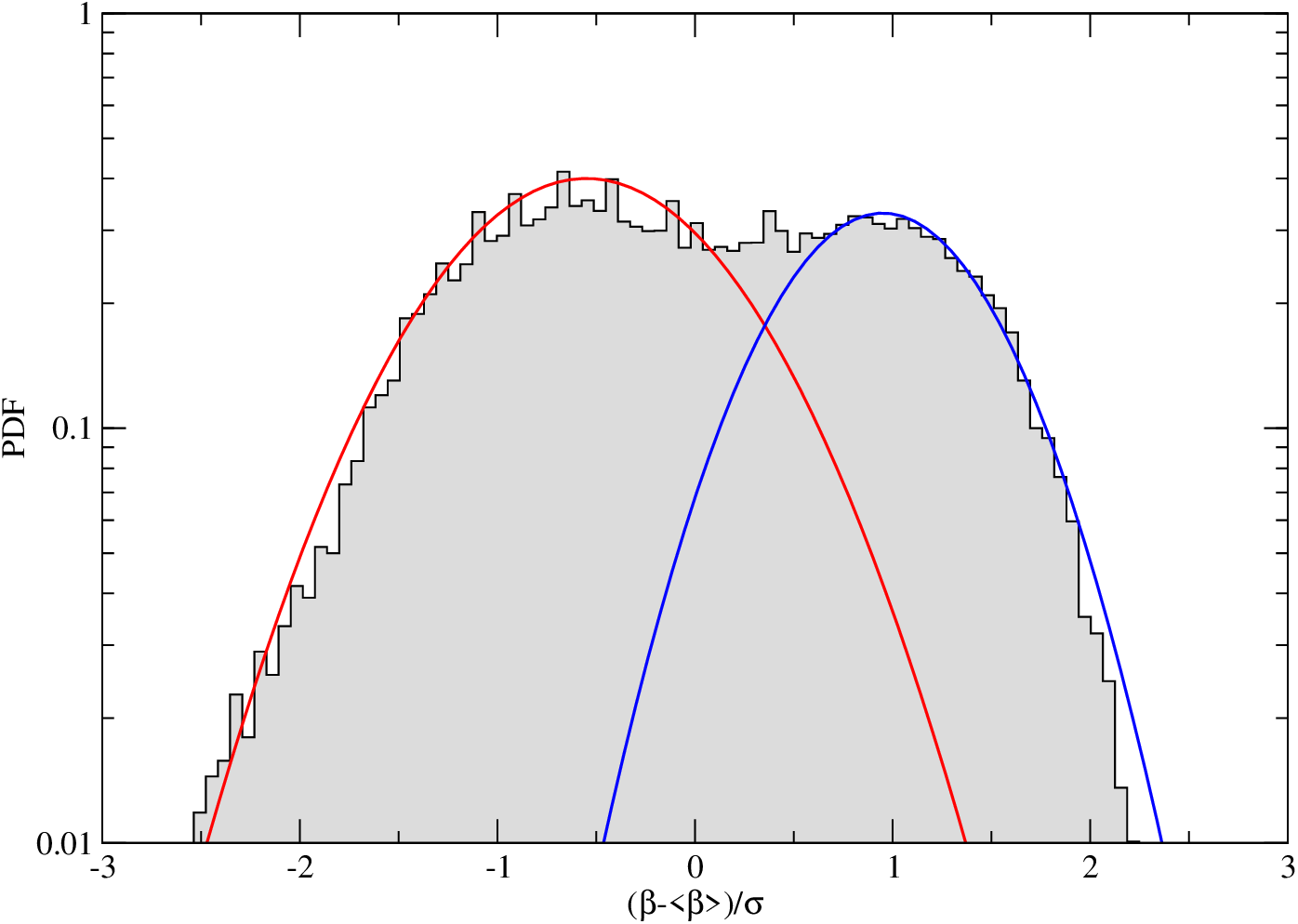, angle=0., width=11cm}
\caption{Distribution of the daily measured inverse mean temperature in Sydney for 1910-2011}

\vspace{2cm}

\epsfig {file=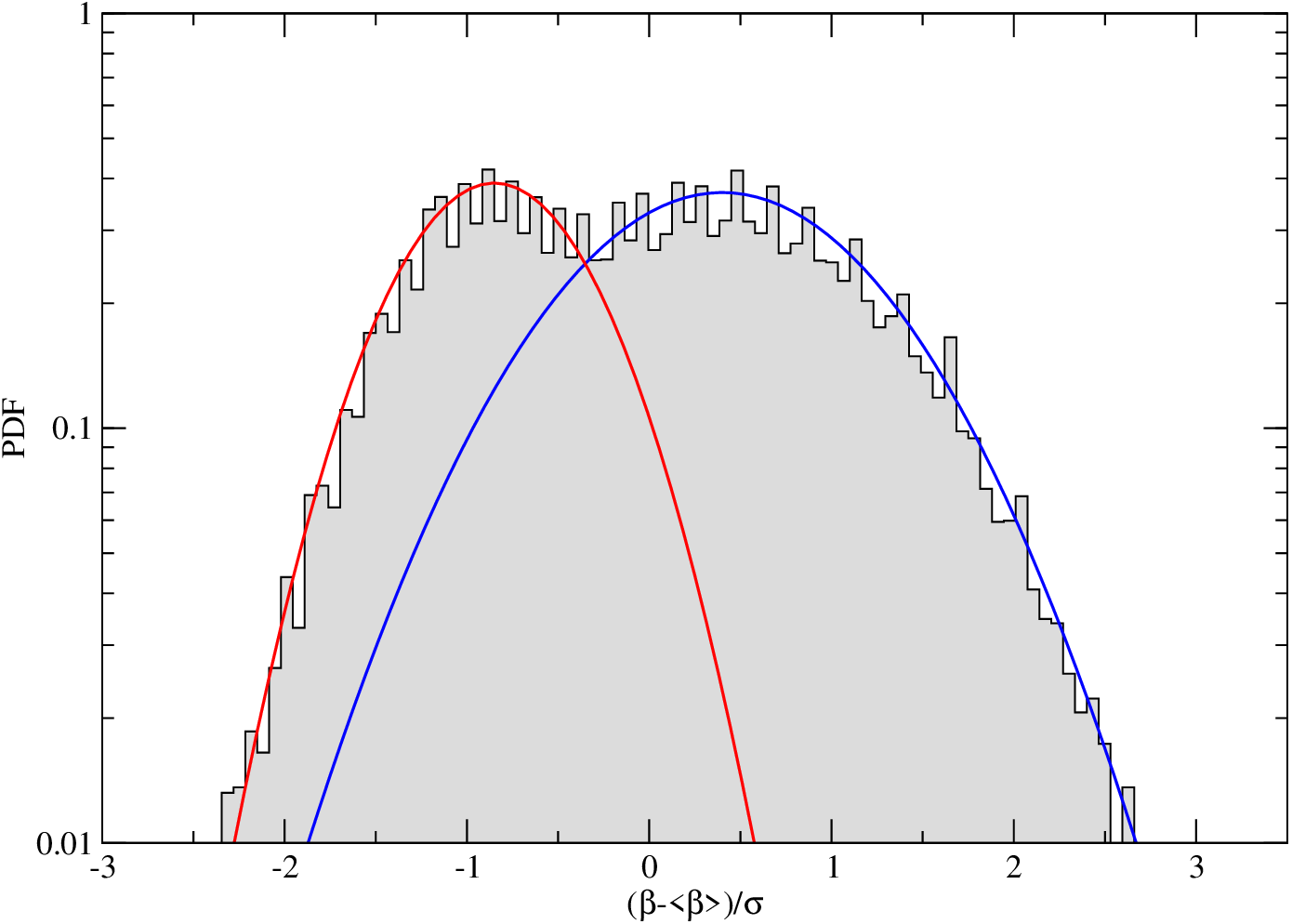, angle=0., width=11cm}
\caption{Distribution of the daily measured inverse mean temperature in Central England (Lancashire, London and Bristol) for 1910-2011}
\end{figure}


\begin{figure} [ht]
\epsfig {file=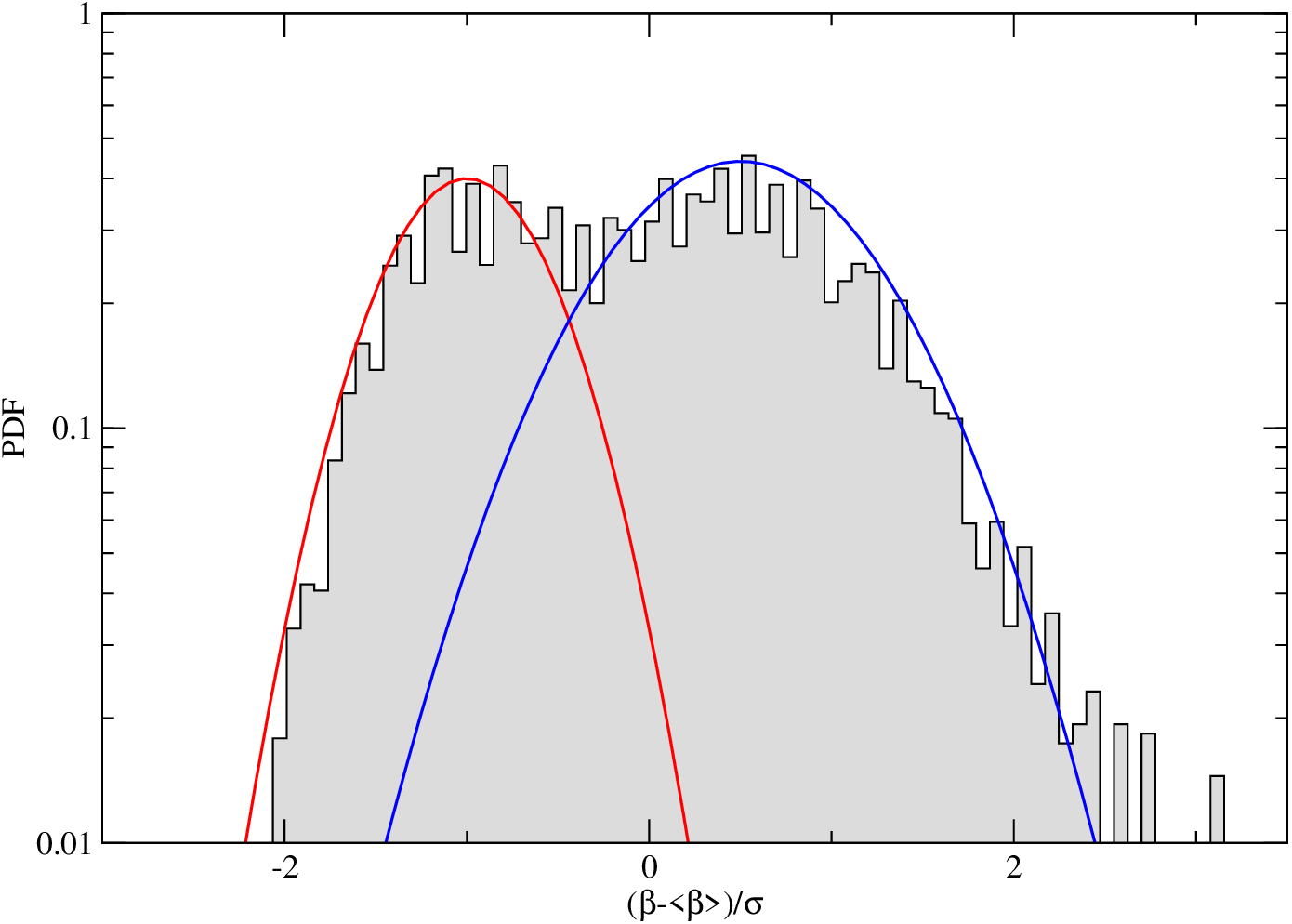, angle=0., width=11cm}
\caption{Distribution of the daily measured inverse mean temperature in Vancouver for 1937-2011}

\vspace{2cm}

\epsfig {file=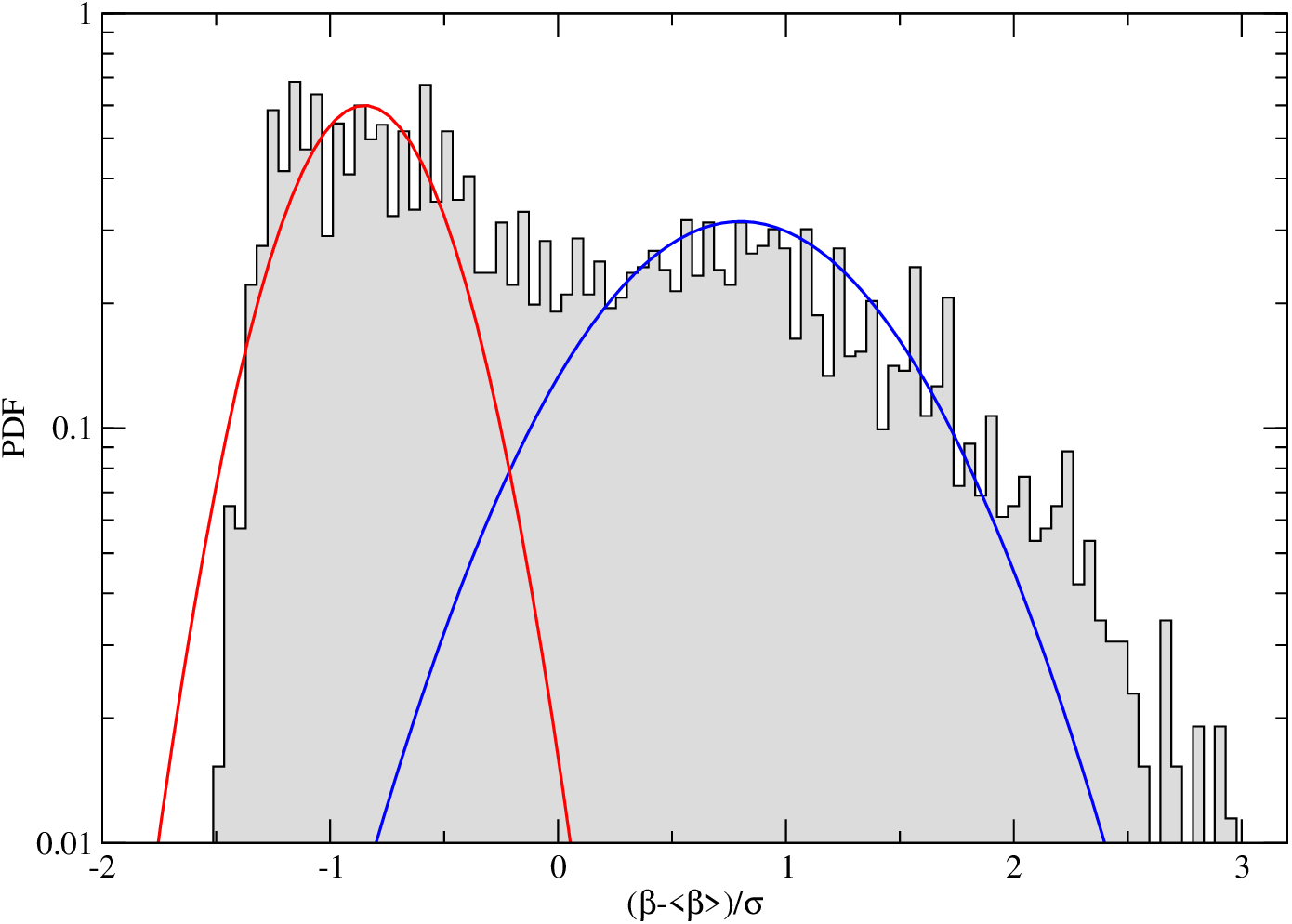, angle=0., width=11cm}
\caption{Distribution of the daily measured inverse mean temperature in Hong Kong for 1997-2011}
\end{figure}

\clearpage

\begin{figure} [ht]
\epsfig {file=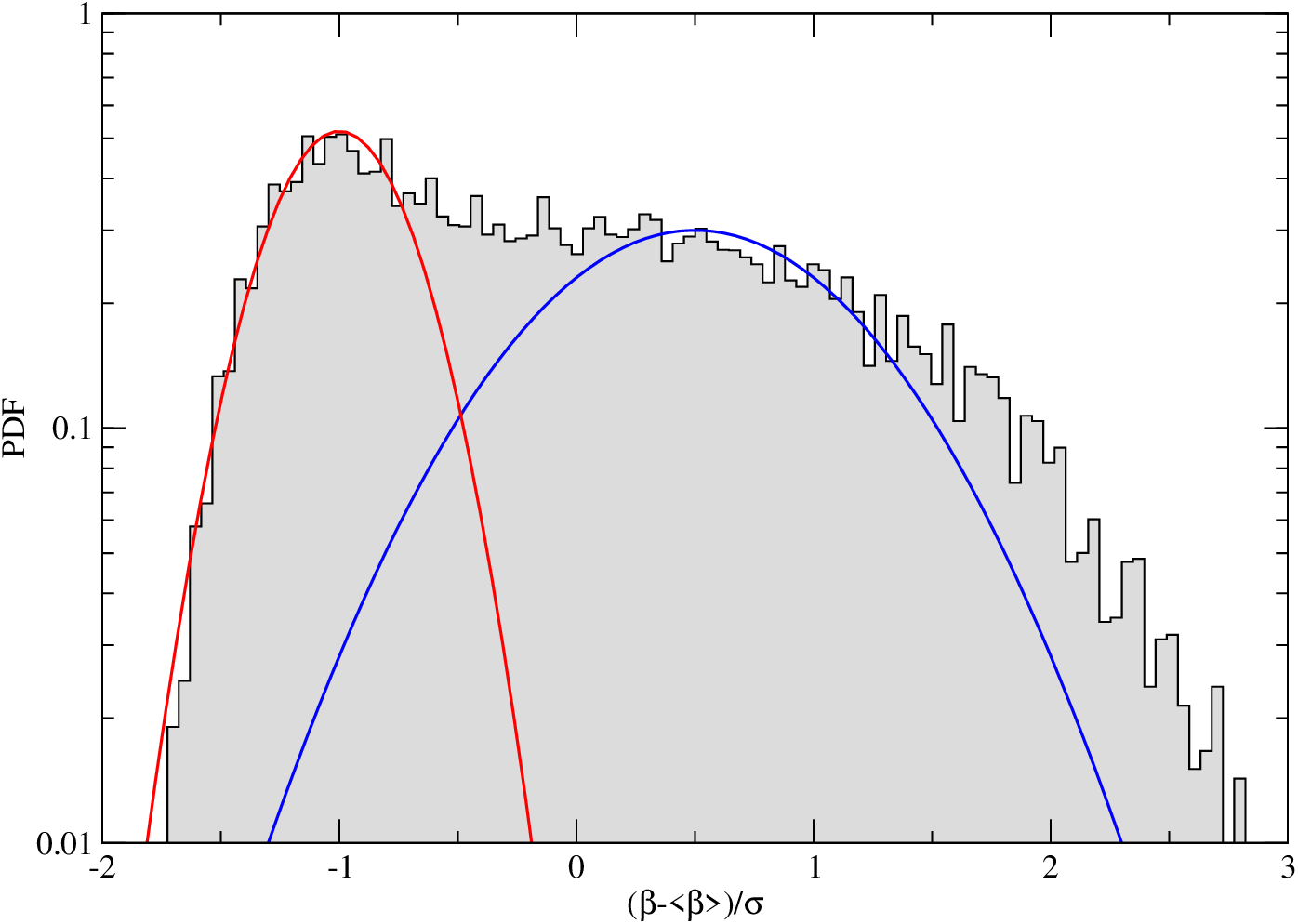, angle=0., width=11cm}
\caption{Distribution of the daily measured inverse mean temperature in Ottawa for 1939-2011}

\vspace{2cm}


\epsfig {file=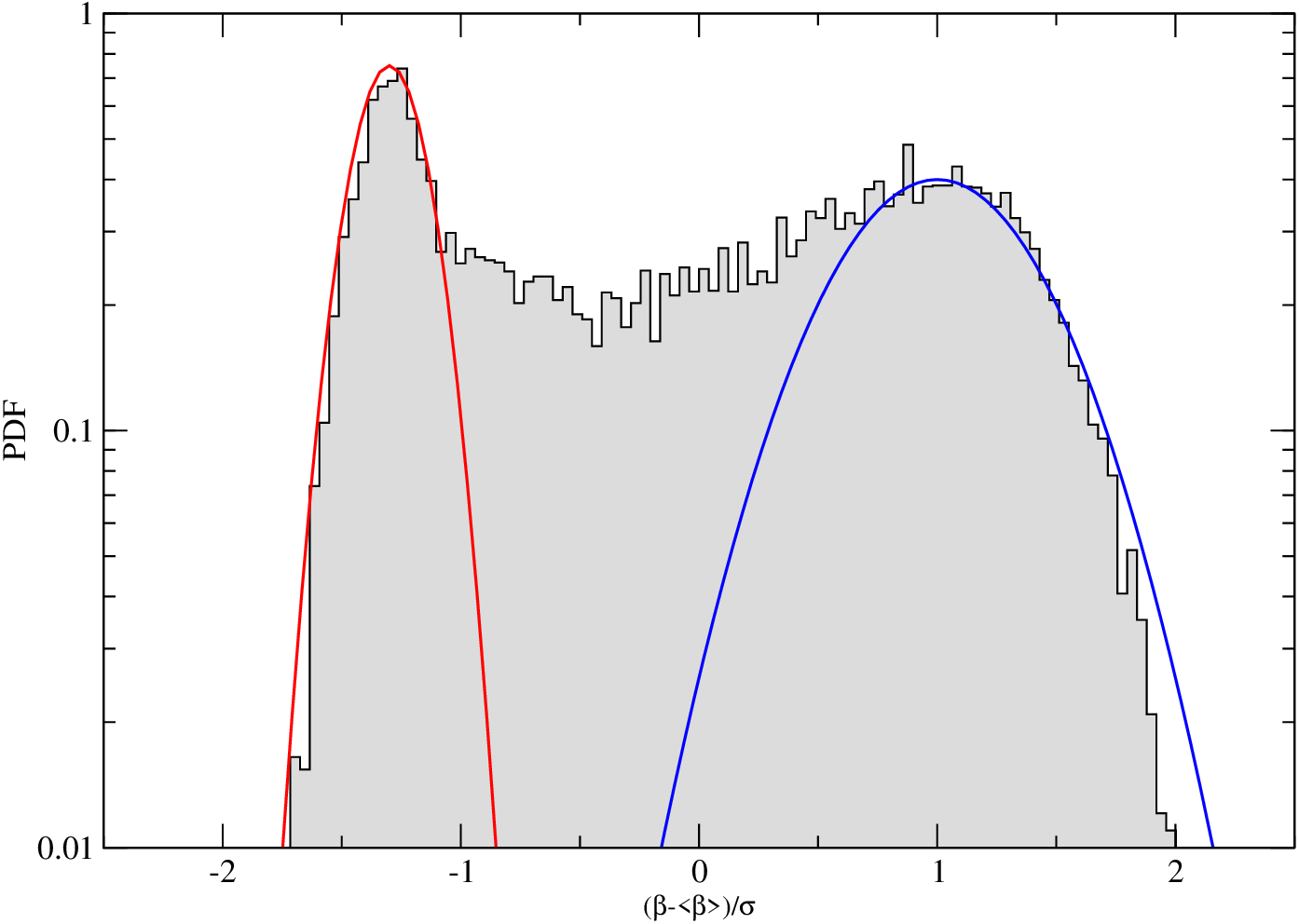, angle=0., width=11cm}
\caption{Distribution of the daily measured inverse mean temperature in Eureka for 1951-2011}
\end{figure}


One may try to fit the data by other functional forms than a superposition of two Gaussians.
For example, the histogram of daily measured inverse mean temperature in Dubai 1974-2011
(Fig.~31) is well fitted by an exponential
$e^{-V(\beta)}$ of a double-well potential $V(x)\sim (C_2x^2+C_3x^3+C_4x^4)$.
Generally, however, there is no simple analytic expression for the environmentally relevant inverse temperature
distributions.

\vspace{0.5cm}


\begin{figure} [ht]
\epsfig {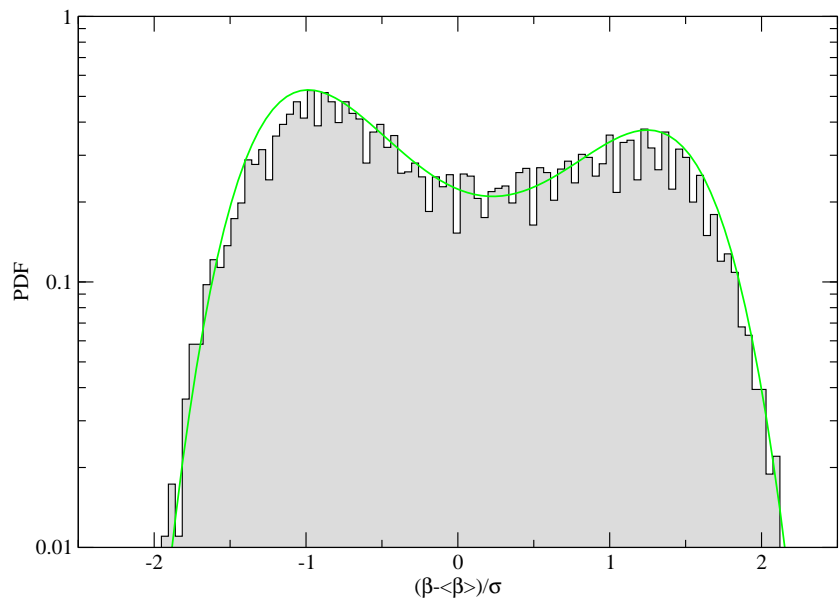}
\caption{Fit with the exponential of a double-well potential to the histogram of the daily measured inverse mean temperature in Dubai 1974-2011}
\end{figure}

\section{Interpretation of results in terms of K\"oppen-Geiger climate classification system}

Our superstatistical distributions
 are consistent with the K\"oppen-Geiger climate classification system \cite{koppen1,koppen2,koppen3} which is one of the most widely used climate classification systems.
In the following we will briefly review this scheme and then interpret our long-term superstatistical
distributions in the K\"oppen-Geiger context. Of course, so far our investigation only involves temperature
distributions. In future work, we intend to take into account rainfall data as well \cite{porporato,kilsby},
which can also be modeled using superstatistical techniques.

There are basically 5 different climatic groups in the K\"oppen-Geiger scheme:
Group A Tropical climates, group B Arid climates, group C Temperate climates, group D Cold climates and group E Polar climates.
We have chosen at least one example in each of these groups:
A: Darwin, B: Santa Fe, Dubai C: Sydney, Central England, Hongkong D: Ottawa, E: Eureka.






A second and third letter give more specific information within each of the groups.
In our examples, we have collected data for the following climate types:

Aw (Tropical-Savannah): Darwin,

BSk (Arid-Steppe-Cold): Santa Fe,

BWh (Arid-Desert-Hot): Dubai,

Cfa (Temperate-Without dry season-Hot summer):  Sydney,

Cfb (Temperate-Without dry season-Warm summer): Central England and Vancouver,

Cwa (Temperate-Dry winter-Hot summer): Hong Kong,

Dfb (Cold-Without dry season-Warm summer): Ottawa,

ET(Polar-Tundra): Eureka.

We summarize below short characterizations of these subtypes:

Aw (Tropical-Savannah) subtype is characterized by the coldest month which has a temperature
 of 18$^{\circ} C$ or more. There is no rainforest and the precipitation of the driest month is less than 100 - mean annual precipitation /25.

BSk (Arid-Steppe-Cold) subtype is characterized by mean annual precipitation  equal or bigger than five times and less than ten times of the precipitation threshold.
The mean annual temperature is less than 18$^{\circ} C$.

BWh (Arid-Desert-Hot) subtype is characterized by mean annual precipitation equal or less than five times of precipitation threshold.
The mean annual temperature is equal or bigger than 18$^{\circ} C$.

Cfa (Temperate-Without dry season-Hot summer) subtype is characterized by
the temperature of the hottest month being equal and bigger than 22$^{\circ} C$ and the temperature of the coldest month is between 0$^{\circ} C$ and 18 $^{\circ} C$. There is no dry season for summer and winter.

Cfb (Temperate-Without dry season-Warm summer) subtype is characterized as follows:
The temperature of the hottest month is bigger than 10$^{\circ} C$ and
the temperature of the coldest month is between 0$^{\circ} C$ and 18$^{\circ} C$.
There is no dry season for summer and winter.There is no hot summer.
The number of months where the temperature is above 10$^{\circ} C$ is equal or bigger than 4.

Cwa(Temperate - Dry winter - Hot summer):
The temperature of the hottest month is equal or bigger than 22$^{\circ} C$.
The temperature of the coldest month is between 0$^{\circ} C$ and 18$^{\circ} C$.
Precipitation of the driest month in winter is less than one-tenth of precipitation of the wettest month in summer.

Dfb(Cold-Without dry season-Warm summer):
The temperature of the hottest month is larger than 10$^{\circ} C$.
The temperature of the coldest month is equal or less than 0$^{\circ} C$.
There is no dry season for summer and winter and there is no hot summer.
The number of months where the temperature is above 10$^{\circ} C$ is equal or bigger than 4.

ET (Polar-Tundra) subtype is characterized by the temperature of the hottest month being between 0$^{\circ} C$ and 10$^{\circ} C$.




When looking at Fig.~22-30 and Table 1, the observed inverse temperature distributions
are clearly consistent with the climate classifications of the various locations.
The plot for Darwin (Fig.~22) is the only one which has only a single maximum,
occuring at the very high temperature 28$^{\circ} C$,
consistent with the tropical location and the non-existence of seasonal variations.
Still it is interesting to see that the distribution strongly deviates from a Gaussian,
there is a pronounced tail corresponding to large $\beta$.

All other plots exhibit a double-peak structure, corresponding
to summer and winter. However, it is interesting that the variance of
the Gaussians that we use for the fits can be very different.
For example, for Santa Fe (Fig.~23) the summer-Gaussian (on the left) has a
smaller variance than the winter-Gaussian (on the right), whereas for
Dubai (Fig.~25) both Gaussians have about the same variance, and finally for
Sidney (Fig.~26) the summer-Gaussian (on the left) has bigger variance than the winter-Gaussian
(on the right). In this way one may use our plots to provide a
quantitative characterization of K\"oppen-Geiger subtypes, or even a new classification.

The plots of Fig.~26 (Central England) and Fig.~27 (Vancouver) are very similar.
This is to be expected, since both locations fall into the same climate type Cfb.

The locations of the most likely temperatures (i.e. the maxima) observed in
Fig.~28-30 are consistent with typical summer and winter temperatures in Hong Kong,
Ottawa and Eureka. Note that
these temperatures are averaged over
day and night.

As already mentioned,
for tropical locations, such as Darwin (Fig.~22),
the two peaks merge in a single peak,
as expected for regions where there is hardly any difference between
summer and winter temperatures.
On the other hand, for polar locations, the two peaks
are strongly separated,
as shown in Fig.~30. for the example of Eureka.
Here the two Gaussians are very far apart from each other.
We suggest to use the variance of the winter- and summer-Gaussians,
as well as their distance, as useful quantitative measures to characterize
the temperature profile of various climate types.
For example, Sydney seems special since it is the only example where the summer-Gaussian has
significantly higher variance than the winter-Gaussian. Dubai seems special as well,
since both winter and summer-Gaussian seem to have roughly the same variance.
This allows for an alternative fit with the exponential of a double-well potential (Fig.~31).

\section{Superstatistics with double-peaked distributions}

The original concept of superstatistics provided an
approximation method for sharply peaked temperature distributions,
which resulted in a perturbative expansion around the Boltzmann-Gibbs
limiting case \cite{beck-cohen}. Apparently, environmental
superstatistics does not have sharply peaked distributions $f(\beta)$,
but as we saw in the previous section there are broad distributions
that often have two maxima. Hence the effective
Boltzmann factors $B(E)=\int f(\beta) e^{-\beta E}d\beta$ can only
be evaluated numerically. One idea would be to effectively
separate the two maxima and to do a superposition of two
single-peaked superstatistics, one for the summer and one for the winter.
This is particularly useful if the thermodynamic device is supposed to work in one
season with a particular performance, and in the other season with another one.
This concept basically boils down to consider conditional temperature distributions, conditioned
on the season of the year.

Another interesting question one can ask for environmental superstatistics
is the large-energy asymptotics of the generalized Boltzmann factors $B(E)$.
One can apply the techniques
of \cite{touchette} to determine, at least in principle,
the decay of $B(E)$ with $E$ for large $E$. For this the
small-$\beta$ behavior of $f(\beta)$ is relevant. For example,
if $f(\beta)\sim \beta^\gamma$ for small $\beta$, then the generalized
Boltzmann factors decay with a power law for large $E$,
the power law exponent being related to $\gamma$ \cite{touchette}:
\begin{equation}
B(E)\sim E^{-\gamma -1}
\end{equation}
Tsallis $q$-statistics \cite{tsallis1,tsallis-book} is a particular example,
with $q=\langle \beta^2 \rangle/\langle \beta \rangle^2 $\cite{beck-cohen}.
In practice,
for this one would need very precise data of $f(\beta)$ for small
$\beta$, which in our environmental interpretation corresponds to the
statistics of hot summers. The extreme-event statistics of very
hot summers determines the asymptotic large-$E$ behaviour of
the effective thermostatistics of the statistical mechanics device
under consideration.
Thus, to decide on the asymptotic behaviour of environmentally relevant $B(E)$
a systematic investigation of extreme events is needed, namely that of very
hot summers.

\section{Global warming}

As we have already mentioned before, in some of our data one sees a systematic trend which
can be associated with global warming. As it is well-known, global warming
   is the rise in the average temperature of the Earth's atmosphere and of the
   oceans since the late 19th century and its projected continuation. It is primarily caused by increasing concentrations of greenhouse gases produced by human activities such as the burning of fossil fuels and deforestation.
It comes with sea level rises and widespread decreases in snow and ice extent.

Quantitatively,
the Earth's average surface temperature rose by 0.74$\pm$0.18 $^{\circ} C$ over the period 1906-2005
\cite{trenberth,jansen,author}. The rate of warming over the last half of that period was almost double that of the period as a whole (0.13$\pm$0.03$^{\circ} C$ per decade, versus 0.07$\pm$0.02 $^{\circ} C$ per decade). The urban heat island effect is very small and accounts for less than 0.002 $^{\circ} C$ of warming per decade since 1900 \cite{trenberth}.
Temperatures in the lower troposphere have increased between 0.13 and 0.22 $^{\circ} C$ (0.22 and 0.4
$^{\circ} F$) per decade since 1979, according to satellite temperature measurements \cite{jansen}.
Arctic regions are especially vulnerable to the effects of global warming, as has become apparent in the melting sea ice in recent years. Climate models predict much greater warming in the Arctic than the global average \cite{author}.

If sampled over many decades, the precision of some of our data is good enough to reveal the
effects of global warming. As an example we show
in Fig.~32 the average of daily measured mean temperature of every single year in Eureka, Nunavut, Canada from 1951-2011.
This is an arctic region and our plot shows that the average temperature grows by $0.92$ $^{\circ} C$ per decade (Fig.~33).
So in this polar region the warming occurs at a much higher rate than 
the global warming
rate that is averaged over the entire earth.

\vspace{2cm}


\begin{figure} [ht]
\epsfig {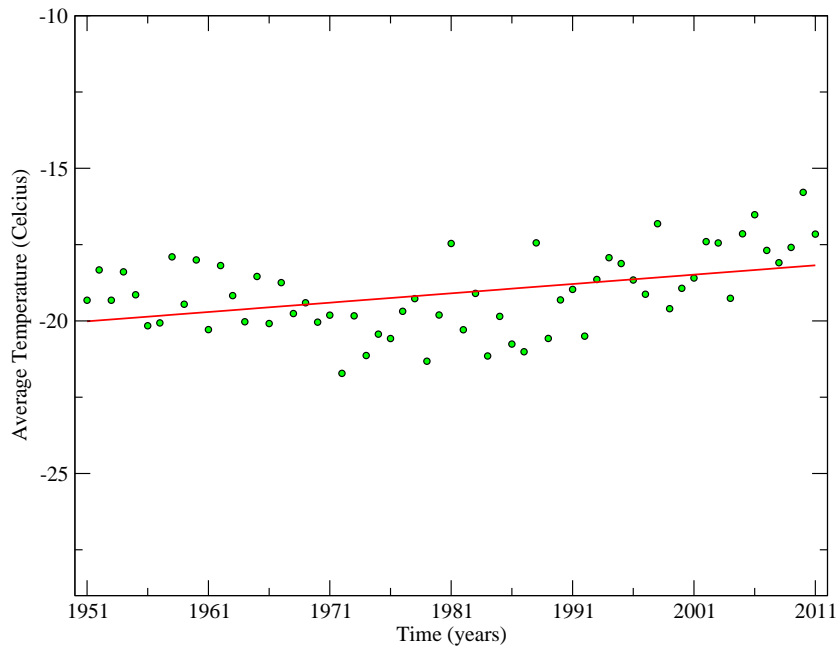}
\caption{Average of daily measured mean temperature of every single year in Eureka 1951-2011 }
\end{figure}

\clearpage

\begin{figure} [ht]
\epsfig {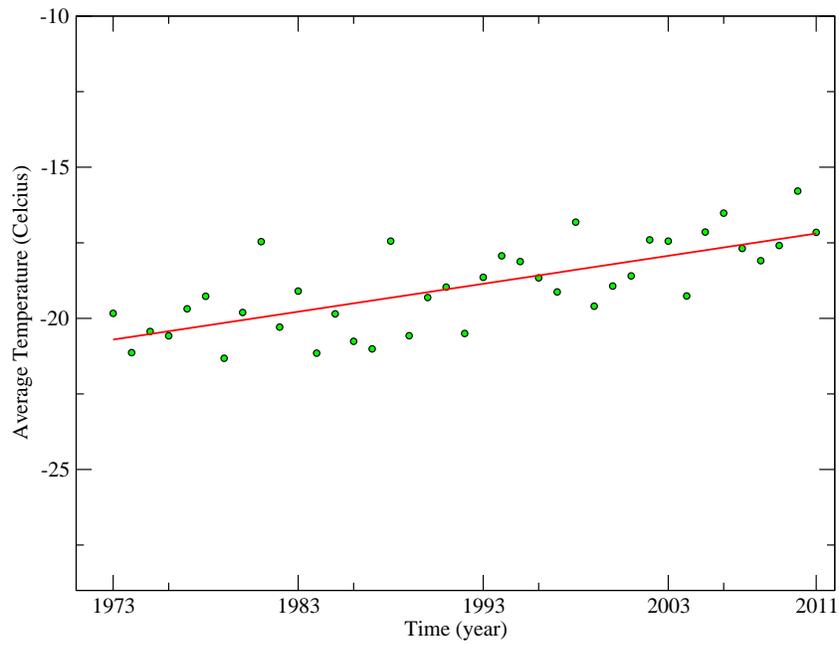}
\caption{Average of daily measured mean temperature of every single year in Eureka 1973-2011 }
\end{figure}

 For some of the other locations we also observe a systematic increase in average temperature (Figs.~34-37).
 For Dubai (Fig.~34) the rate is also rather high, $0.63$ $^{\circ} C$ per decade. For Sydney (Fig.~35),
 Vancouver (Fig.~36), and Ottawa (Fig.~37) one has rates comparable to the average global warming rate,
 namely $0.13$, $0.11$ and $0.18$ $^{\circ} C$ per decade, respectively.
It is in the remit of the superstatistics approach that it can
very well model both stochastic effects as well as systematic drifts, as long as there is a clear separation of
time scales.


\vspace{2cm}

\begin{figure} [ht]
\epsfig {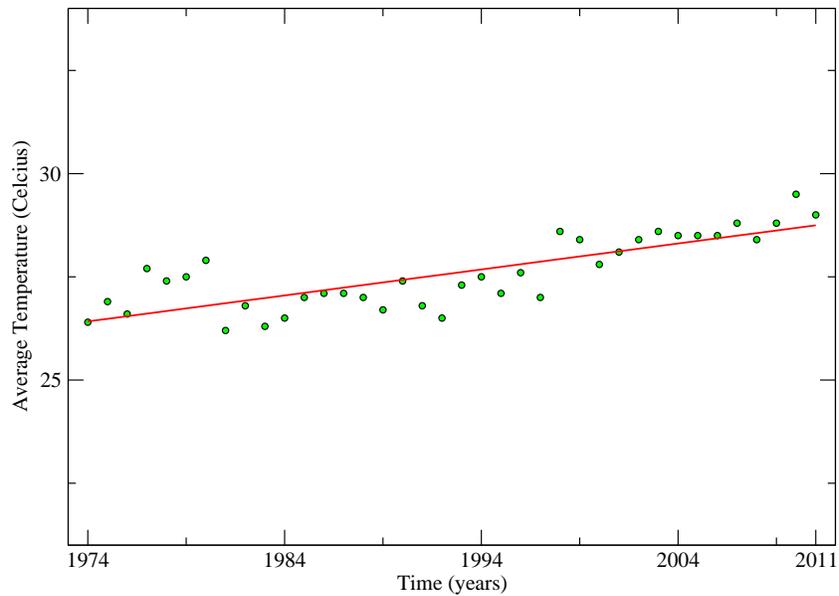}
\caption{Average of daily measured mean temperature of every single year in Dubai 1974-2011 }
\end{figure}

\begin{figure} [ht]
\epsfig {file=sydney-global.eps, angle=0., width=11cm}
\caption{Average of daily measured mean temperature of every single year in Sydney 1910-2011 }

\vspace{2cm}


\epsfig {file=vancouver-global.eps, angle=0., width=11cm}
\caption{Average of daily measured mean temperature of every single year in Vancouver 1937-2011 }
\end{figure}

\clearpage


\begin{figure} [ht]
\epsfig {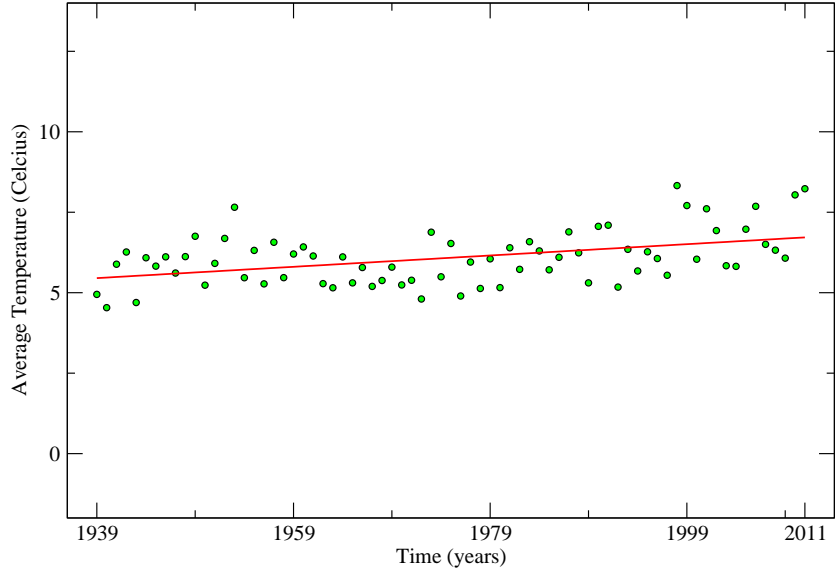}
\caption{Average of daily measured mean temperature of every single year in Ottawa 1939-2011 }
\end{figure}

{In the following we illustrate that global warming (i.e., a slight increase in
average temperature) has an even  more dramatic effect if the fluctuations of temperature
at the various spatial locations
are taken into account. To see this, let us consider a simple toy model, namely that
of a $\chi$-square distribution of inverse temperature of degree $n$, as given by
\begin{equation}
 f(\beta)=\frac{1}{\Gamma (\frac{n}{2})}\left(\frac{n}{2\beta_0}\right)^{\frac{n}{2}} \beta^{\frac{n}{2}-1} e^{-\frac{n\beta}{2\beta_0}}.
\label{eq11}
\end{equation}
Near the maximum, this distribution is close to a Gaussian distribution, for which we had
evidence in section 2, provided we condition the observed temperatures onto a particular month of the year.
For this toy model the average of the fluctuating $\beta$ is given by
\begin{equation}
 \langle \beta\rangle =\int_0^\infty \beta f(\beta)=\beta_0
\end{equation}
and the variance by
\begin{equation}
 \langle \beta^2\rangle -\beta_0^2=\frac{2}{n}\beta_0^2
\label{fran}
\end{equation}
Let us now couple this environmental temperature bath to a local system, which for simplicity we simply take
to be an ideal gas, although more generally we may think of more complicated
ecosystems influenced by the temperature fluctuations of its environment. The kinetic energy of a test particle is given by $E=\frac{1}{2} v^2$,
and for a given inverse temperature the velocity $v$ is distributed according to the Gaussian distribution
\begin{equation}
p(v|\beta)=\sqrt{\frac{\beta}{2\pi}}e^{-\frac{1}{2}\beta v^2}.
 \label{eq10}
\end{equation}
             Taking into account the environmental
             temperature fluctuations, the long-term probability density to observe the velocity
 $v$ of the test particle is given by the marginal probability $p(v)$ as
\begin{equation}
 p(v)=\!\! \! \int_0^\infty f(\beta) p(v|\beta) d\beta\\
\label{eq14}
\end{equation}
which in our toy example case can be evaluated as
\begin{equation}
 p(v)
=\frac{\Gamma(\frac{n}{2}+\frac{1}{2})}{\Gamma(\frac{n}{2})} \left(\frac{\beta_0}{\pi n}\right)^\frac{1}{2} \frac{1}{\left(1+\frac{\beta_0}{n} v^2\right)^{\frac{n}{2}+\frac{1}{2}}}.
\label{eq16}
\end{equation}
Indeed we obtain $q$-statistics \cite{tsallis1, tsallis-book} where the entropic parameter $q$ is related
to the number of degrees of freedom influencing the $\beta$-variable by $q-1=\frac{2}{n+1}$
\cite{prl2001}. More generally, also for other distributions $f(\beta)$, one may define a parameter
$q$ measuring the strength of environmental inverse temperature fluctuations by \cite{beck-cohen}
\begin{equation}
q = \frac{\langle \beta^2 \rangle}{\langle \beta \rangle^2}. \label{fran2}
\end{equation}
If there are no temperature fluctuations, i.e.\ $\beta=\beta_0=const$, then $q=1$, otherwise $q>1$.

We are now in a position to quantitatively calculate the response
to global warming, i.e. a slight decrease in the average inverse temperature $\beta_0$
for this superstatistical model system. Consider the mean kinetic energy associated
with velocity $v$ of a test particle of mass 1,
given by
\begin{equation}
\frac{1}{2} \bar{v^2} =\frac{1}{2}\int_{-\infty}^{+\infty} p(v)v^2dv,
\end{equation}
where $\bar{ \cdots}$ denotes the expectation with respect to the marginal distribution $p(v)$.
In an environmental setting, within our toy model, we may think of this as the mean kinetic energy intensity
of particle movements, making up e.g.\ storms, in an environment where (inverse) temperature is distributed according to
a $\chi$-square distribution. For $q$-statistics one can explicitly calculate the above integral as
\begin{equation}
\frac{1}{2} \bar{ v^2} = T_0 \frac{3-q}{5-3q},
\end{equation}
where $T_0=1/\beta_0$ is the mean temperature.

We now see that the effects of global warming
are more dramatic if temperature fluctuations ($q >1$) are taken into account: The change
with global warming (i.e., a slight increase of $T_0$) is given by
\begin{equation}
\frac{1}{2} \frac{d}{d T_0} \bar{v^2 } = \frac{3-q}{5-3q}. \label{tom}
\end{equation}
If $q=1$ (no temperature fluctuations), the quantity 
on the right hand side is just 1, which
corresponds to the mean field response to global warming if no temperature fluctuations are
taken into account. However, as shown in this paper it is important to take into
account the distribution of local temperature, which
creates, according to eq.~(\ref{fran}) or (\ref{fran2}) an effective 
 $q>1$. For $q>1$ the term on the right-hand side of eq.~(\ref{tom}) is larger than 1,
thus meaning a more dramatic response effect to global warming. The expected kinetic energy
of a test particle grows stronger than in the case where there are
no temperature fluctuations, and thus ultimately we also expect that storm intensity
or other environmentally interesting observables
will respond more strongly to tiny increases of $T_0$.


\section{Conclusion}

In this paper we have analyzed in detail the superstatistical distributions $f(\beta)$
relevant for a generalized statistical mechanics
description of complex systems that are coupled to a changing temperature environment
on planet earth. These local complex systems may include thermodynamic devices working outdoors, but in a more
general setting can also include local ecosystems and other environmentally relevant subunits.
We looked at local inverse temperature
distributions for numerous examples of spatial locations. If seasonal periodicity is not
removed from the data, typical distributions have a double-peak structure, with details
of the two peaks (their intensity, variance, and intermediate connection) dependent on
the local climate. The double-peak structure implies that conventional type of superstatistics \cite{beck-cohen}
for single-peaked distributions is not applicable. If data are restricted to particular
months, thus eliminating seasonal variations, one obtains single-peaked distributions, which in reasonably good approximation
are Gaussian, at least in the vicinity of the
maximum and for most examples of spatial locations we have chosen.
However, for some locations (e.g. Vancouver, Darwin) we do see anomalous tails, in line with
recent observations reported in
\cite{ruff}, though in our case this seems to be more the exception than the rule.
A planned future project is to
analyse the tails in temperature
distributions (as well as precipitation statistics) in more detail at more locations, taking into account
known techniques from extreme event statistics \cite{kantz}.

On time scales of several decades our data clearly show the effect of global warming, with the rate
of average increase
in temperature depending very much on spatial location. Some examples of locations (Eureka, Dubai)
were found where the rate of increase of average temperature is significantly higher than the global average.
When looking at the response effects to global warming
it is very important to take into account local temperature distributions and thus to proceed
to a superstatistical description.
We showed (for a simple toy model system)
that the effect of global warming on kinetic energy expectations (of moving particles)
is indeed stronger if temperature distributions at the various spatial locations (rather than constant
temperature) are taken into account.}

\section{Acknowledgement}
G.Cigdem Yalcin was supported by the Scientific Research Projects Coordination Unit of Istanbul University with project number 7441. She gratefully acknowledges the hospitality of Queen Mary University of London, School of Mathematical Sciences, where this work was carried out. Christian Beck's research
is supported by the EPSRC grant `Flood MEMORY'.


\begin{thebibliography}{99}
\bibitem{beck-cohen} C. Beck and E.G.D. Cohen,  Physica A {\bf 322}, 267 (2003)
\bibitem{swinney} C. Beck, E.G.D. Cohen, and H.L. Swinney, Phys. Rev. E {\bf 72}, 056133 (2005)
\bibitem{touchette} H. Touchette and C. Beck,
Phys. Rev. E {\bf 71}, 016131 (2005)
\bibitem{souza} C. Tsallis and A.M.C. Souza,
Phys. Rev. E {\bf 67}, 026106 (2003)
\bibitem{jizba} P. Jizba, H. Kleinert,
Phys. Rev. E {\bf 78}, 031122 (2008)
\bibitem{chavanis} P.-H. Chavanis,
Physica A {\bf 359}, 177 (2006)
\bibitem{frank} S.A. Frank and D.E. Smith, Entropy {\bf 12}, 289 (2010)
\bibitem{celia} C. Anteneodo and S.M. Duarte Queiros, J. Stat. Mech. P10023 (2009)
\bibitem{straeten} E. Van der Straeten and C. Beck, Phys. Rev. E {\bf 80}, 036108 (2009)
\bibitem{hanel} R. Hanel, S. Thurner, and M. Gell-Mann,
PNAS {\bf 108}, 6390 (2011)
\bibitem{briggs} K. Briggs, C. Beck,
Physica A {\bf 378}, 498 (2007)
\bibitem{prl} C. Beck,
Phys. Rev. Lett. {\bf 98}, 064502 (2007)
\bibitem{chen} L. Leon Chen, C. Beck,
Physica A {\bf 387}, 3162 (2008)
\bibitem{abul-magd} A.Y. Abul-Magd, G. Akemann, P. Vivo, J. Phys. A Math. Theor. {\bf 42}, 175207 (2009)
\bibitem{daniels} K.E. Daniels, C. Beck, and E. Bodenschatz,
Physica D {\bf 193}, 208 (2004)
\bibitem{cosmic} C. Beck,
Physica A {\bf 331}, 173 (2004)
\bibitem{rapisarda} S. Rizzo and A. Rapisarda,
AIP Conf. Proc. {\bf 742}, 176 (2004)
\bibitem{soby} D.N. Sobyanin, Phys. Rev. E {\bf 24}, 051128 (2011)
\bibitem{dixit} P.D. Dixit, arXiv:1210.3015
\bibitem{tsallis1} C. Tsallis, J. Stat. Phys. {\bf 52}, 479 (1988)
\bibitem{tsallis-book} C. Tsallis, Introduction to
Nonextensive Statistical Mechanics, Springer, 2009
\bibitem{prl2001} C. Beck, Phys. Rev. Lett.  {\bf 87}, 180601 (2001)
\bibitem{lu1} V. Lucarini, T. Nanni, A. Speranza,  Nuovo Cimento C {\bf 27}, 285 (2004)
\bibitem{lu2} M.R. Attolini, S. Cecccinni, M. Galli, Nuovo Cimento C {\bf 7}, 245 (1984)
\bibitem{lu3} K.Y. Vinnikov, A. Robock, N.C. Grody, A. Basist, Geophys. Res. Lett. {\bf 31}, L06205 (2003)
\bibitem{porporato} A. Porporato, G. Vico, P.A. Fay, Geophys. Res. Lett. {\bf 33}, L15402 (2006)
\bibitem{kilsby} M.R. Jones, H.J. Fowler, C.G. Kilsby, S. Blenkinsop, Int. J. Climatology (2012),
DOI: 10.1002/joc.3503
\bibitem{australia} Australian Government Bureau of Meteorology www.bom.gov.au
\bibitem{ncdc} National Oceanic and Atmospheric Administration, National Climatic Data Center www.ncdc.noaa.gov
\bibitem{dubai} Dubai Meteorology Ofice  www.dia.ae
\bibitem{uk1} D.E. Parker, T.P. Legg, and C.K. Folland.  A new daily Central England Temperature Series, 1772-1991. Int. J. Clim., Vol 12, 317-342 (1992)
\bibitem{uk2} United Kingdom National Weather Service www.metoffice.gov.uk
\bibitem{canada} Enviroment Canada Weather Office www.weatheroffice.gc.ca
\bibitem{hongkong}Hong Kong Observatory, The Government of the Hong Kong Special Administrative Region  www.hko.gov.hk


\bibitem{koppen1} W. K\"oppen, Meteorologische Zeitschrift, {\bf 20}, 3, 351 (2011)
\bibitem{koppen2} M. Kottek, J. Grieser, C. Beck, B. Rudolf, F. Rubel, Meteorologische Zeitschrift, {\bf 15}, 3, 259 (2006)
\bibitem{koppen3} M. C. Peel, B. L. Finlayson, T. A. McMahon, Hydrol. Earth Syst. Sci., {\bf 11}, 1633 (2007)


\bibitem{trenberth} Trenberth et al., Ch. 3, Observations: Atmospheric Surface and Climate Change, Section 3.2.2.2: Urban Heat Islands and Land Use Effects, p. 244, in IPCC AR4 WG1 2007.

\bibitem{jansen} Jansen et al., Ch. 6, Palaeoclimate, Section 6.6.1.1: What Do Reconstructions Based on Palaeoclimatic Proxies Show?, pp. 466-478, in IPCC AR4 WG1 2007.

\bibitem{author} S.J. Hassol, Impacts of a warming Arctic: Arctic Climate Impact Assessment. Cambridge, UK: Cambridge University Press. February 2005. doi:10.2277/0521617782. ISBN 0-521-61778-2.

{
\bibitem{ghil} R. Vautard, P. Yiou, M. Ghil, Physica D {\bf 58}, 95 (1992)

\bibitem{ghil2} M. Ghil and R. Vautard, Nature {\bf 350}, 324 (1991)

\bibitem{allen} M.R. Allen, P.L. Read, L.A. Smith, Nature {\bf 355}, 686 (1992)

\bibitem{porporato2} G. Katul, A. Porporato, R. Oren,
Annu. Rev. Ecol. Evol. Syst. {\bf 38}, 767 (2007)

\bibitem{kantz} M.S. Santhanam and H. Kantz, Phys. Rev. E {\bf 78},
051113 (2008)

\bibitem{ruff} T.W. Ruff and J.D. Neelin, Geophys. Res. Lett. {\bf 39},
L04704 (2012)
}

\end{thebibliography}
\end{document}